\documentclass[preprint,showpacs,preprintnumbers,amsmath,amssymb,floatfix]{revtex4}

\usepackage{graphicx}% Include figure files
\usepackage{dcolumn}% Align table columns on decimal point
\usepackage{bm}% bold math
\usepackage{epsfig}
\begin{document}

\preprint{DESY~09--008, MZ--TH/09--03, LPSC 09--17
\hspace{6.0cm} ISSN 0418-9833}
%\preprint{MZ--TH/09--03, LPSC 08--??\hspace{13.cm}}
%\preprint{January 2009\hspace{14.9cm}}

\title{Open charm hadroproduction and the charm content of the proton}

\author{Bernd A. Kniehl}
\email{kniehl@desy.de}
\author{Gustav Kramer}
\email{gustav.kramer@desy.de}
\affiliation{{II.} Institut f\"ur Theoretische Physik,
Universit\"at Hamburg, Luruper Chaussee 149, 22761 Hamburg, Germany}
\author{Ingo Schienbein}
\email{schien@lpsc.in2p3.fr}
\affiliation{Laboratoire de Physique Subatomique et de Cosmologie,
Universit\'e Joseph Fourier Grenoble 1, CNRS/IN2P3, Institut National
Polytechnique de Grenoble, 53 avenue des Martyrs, 38026 Grenoble, France}
\author{Hubert Spiesberger}
\email{hspiesb@thep.physik.uni-mainz.de}
\affiliation{Institut f\"ur Physik, Johannes-Gutenberg-Universit\"at,
Staudinger Weg 7, 55099 Mainz, Germany}

\date{\today}

\begin{abstract}
We advocate charmed-hadron inclusive hadroproduction as a laboratory to probe
intrinsic charm (IC) inside the colliding hadrons.
Working at next-to-leading order in the general-mass variable-flavor-number
scheme endowed with non-perturbative fragmentation functions recently
extracted from a global fit to $e^+e^-$ annihilation data from KEKB, CESR, and
LEP1, we first assess the sensitivity of Tevatron data of $D^0$, $D^+$, and
$D^{*+}$ inclusive production to the IC parameterizations provided by Pumplin
{\it et al.}
We then argue that similar data from $pp$ collisions at RHIC would have the
potential to discriminate between different IC models provided they reach out
to sufficiently large values of transverse momentum.
\end{abstract}

\maketitle

\section{Introduction}

Recently, the inclusive production of charmed hadrons ($X_c)$ at hadron
colliders has been the subject of extensive experimental and theoretical
studies.
The CDF Collaboration measured the differential cross section
$\mathrm{d}\sigma/\mathrm{d}p_T$ for the inclusive production of $D^0$, $D^+$,
$D^{*+}$, and $D_s^+$ mesons (and their antiparticles) in $p\overline{p}$
collisions at the Fermilab Tevatron (run II), with center-of-mass energy
$\sqrt{s}=1.96$~TeV, as a function of the transverse momentum ($p_T$) in the
central rapidity ($y$) region \cite{Acosta:2003ax}.
The PHENIX Collaboration measured non-photonic electron production through
charm and bottom decays in $pp$, $d$Au, and AuAu collisions at the BNL
Relativistic Heavy Ion Collider (RHIC) with $\sqrt{s}=200$~GeV as a function
of $p_T$ \cite{Adare:2006hc}.
The STAR Collaboration at RHIC presented mid-rapidity open charm spectra from
direct reconstruction of $D^0/\overline{D}^0\to K^\mp\pi^\pm$ decays in $d$Au
collisions and indirect electron-positron measurements via charm semileptonic
decays in $pp$ and $d$Au collisions at $\sqrt{s}=200$~GeV \cite{Adams:2004fc}.
Recently, they also reported results on non-photonic electron production in
$pp$, $d$Au, and AuAu collisions at $\sqrt{s}=200$~GeV \cite{Abelev:2006db}.
Unfortunately, these RHIC data only cover a very limited small-$p_T$ range,
where theoretical predictions based on perturbative QCD are difficult.

On the theoretical side, the cross sections for the inclusive production of
$X_c$ mesons can be obtained as convolutions of universal parton distribution
functions (PDFs) and universal fragmentation functions (FFs) with
perturbatively calculable hard-scattering cross sections.
The non-perturbative input in form of PDFs and FFs must be known from fits to
other processes.
The universality of the PDFs and FFs guarantees unique predictions for the
cross section of the inclusive production of heavy-flavored hadrons.
The results of such calculations for $X_c$ production at the Tevatron have
been presented recently by us in Ref.~\cite{Kniehl:2005st} and compared to the
CDF data \cite{Acosta:2003ax}.
For all four meson species, $D^0$, $D^+$, $D^{*+}$, and $D_s^+$,  we found
good agreement with the data in the sense that the experimental and
theoretical errors overlap.
For the $D^0$, $D^{*+}$, and $D_s^+$ mesons, many of the central data points
fall into the theoretical error band due to scale variation.
Only the central data points for the $D^+$ mesons lie somewhat above it.
For the $D^0$, $D^+$, and $D^{*+}$ mesons, the central data points tend to
overshoot the central theoretical prediction by a factor of about 1.5 at
the lower end of the considered $p_T$ range.
With the exception of the $D_s^+$ case, the experimental results are gathered
on the upper site of the theoretical error band, corresponding to a small
value of the renormalization scale $\mu_R$ and large values of the
factorization scales $\mu_F$ and $\mu_F^\prime$, related to the initial and
final states, respectively.

In our analysis \cite{Kniehl:2005st}, we employed the general-mass
variable-flavor-number scheme (GM-VFNS), which combines the zero-mass
variable-flavor-number scheme (ZM-VFNS) and the fixed-flavor-number scheme
(FFNS).
The GM-VFNS is close to the ZM-VFNS, but retains $m^2/p_T^2$ power terms in
the hard-scattering cross sections.
Here, $m$ stands for the mass of the charm quark.
The calculational details for hadron-hadron collisions were elaborated in our
previous works \cite{Kniehl:2004fy}.
The characteristic feature of the GM-VFNS is that the charm quark is also
treated as an incoming parton originating from the (anti)proton, leading to
additional contributions besides those from the light $u$, $d$, and $s$ quarks
and the gluon.
This is quite analogous to the ZM-VFNS, but with the important difference that
the power terms are retained in the hard-scattering cross sections.
This allows us to apply the GM-VFNS also in the region of intermediate $p_T$
values, $p_T\agt m$.
In fact, in Ref.~\cite{Kniehl:2005st}, we had $p_T\agt m$.

In Ref.~\cite{Kniehl:2005st}, we included $n_f=4$ active quark flavors and
took the charm-quark mass to be $m = 1.5$~GeV.
The strong-coupling constant $\alpha_s^{(n_f)}(\mu_R)$ was evaluated in
next-to-leading order (NLO) with $\Lambda^{(4)}=328$~MeV, corresponding to
$\alpha_s^{(5)}(m_Z)=0.1181$.
We employed proton PDF set CTEQ6.1M from the Coordinated
Theoretical-Experimental Project on QCD (CTEQ) Collaboration
\cite{Pumplin:2002vw} and the FFs from Ref.~\cite{Kniehl:2006mw}.
The default choice for the renormalization and the initial- and final-state
factorization scales was $\mu_R=\mu_F=\mu_F^\prime=m_T$, where
$m_T=\sqrt{p_T^2+m^2}$ is the transverse mass.

Compared to our previous work \cite{Kniehl:2005st}, we have now much more
reliable FFs for the transitions $u,d,s,c,g\to X_c$ at our disposal.
The FFs used in Ref.~\cite{Kniehl:2005st} were determined by fitting the
fractional-energy spectra of the hadrons $X_c$ measured by the OPAL
Collaboration \cite{Alexander:1996wy,Ackerstaff:1997ki} in $e^+e^-$
annihilation on the $Z$-boson resonance at the CERN LEP1 collider.
These data had rather large experimental errors and the disadvantage of being
at a rather large scale $\mu_F^\prime=M_Z$, far away from the scales of
$X_c$ production in $p\overline{p}$ collisions at the Tevatron, which
typically correspond to $p_T$ values below 25~GeV.
In the meantime, new data on charmed-meson production with much higher
accuracy were presented by the CLEO Collaboraton \cite{Artuso:2004pj} at CESR
and by the Belle Collaboration \cite{Seuster:2005tr} at the KEK collider for
$B$ physics (KEKB).
These data offered us the opportunity to determine the non-perturbative
initial condition of the FFs much more accurately.
The data from CLEO \cite{Artuso:2004pj} and Belle \cite{Seuster:2005tr} are
located much closer to the threshold  $\sqrt{s}=2m$ of the transition
$c\to X_c$ than those from OPAL \cite{Alexander:1996wy,Ackerstaff:1997ki}.
The scale of the CLEO \cite{Artuso:2004pj} and Belle \cite{Seuster:2005tr}
data is set by the c.m.\ energy $\sqrt{s}=10.5$~GeV.
Recently, Kneesch and three of us extracted from these data FFs for $D^0$,
$D^+$, and $D^{*+}$ mesons; the details may be found in
Ref.~\cite{Kneesch:2007ey}.

Another important ingredient for the theoretical description of inclusive
$X_c$ production are the parton distribution functions (PDFs).
Let $f_a(x,\mu_F)$ denote the PDF of parton $a$ inside the proton at momentum
fraction $x$ and factorization scale $\mu_F$.
At short distances, corresponding to large values of $\mu_F$, the scale
dependence is determined by the QCD evolution equations with perturbatively
calculable evolution kernels.
So, the PDFs are fully determined by their functional forms in $x$ specified
at a fixed scale $\mu_F=\mu_0$ provided $\mu_0$ is large enough to be in the
region where perturbative QCD is supposed to the valid.
In most applications, $\mu_0$ is chosen to be of order 1--2~GeV,
which is at the borderline between the short-distance (perturbative) and
long-distance (non-perturbative) regions.
The PDFs of the gluon and the light quarks ($a = g, u, d, s$) are
non-perturbative.
They are obtained phenomenologically through global QCD analyses, in which the
theoretical predictions are compared with a wide range of experimental data on
hard processes \cite{Pumplin:2002vw,Tung:2006tb,Thorne:2006zu}.

The cross sections of inclusive charmed-meson production in the GM-VFNS
depend heavily on the PDF of the charm quark and less on those of the light
quarks.
In the following, we use the short-hand notation $c(x,\mu_F)=f_c(x,\mu_F)$.
In the global analyses of
Refs.~\cite{Pumplin:2002vw,Tung:2006tb,Thorne:2006zu} and many others
\cite{LHAPDF}, the charm quark is considered a parton.
It is characterized by the PDF $c(x,\mu_F)$ that is defined for $\mu_F>m$.
In common global QCD analyses at NLO in the $\overline{\mathrm{MS}}$ scheme,
the charm quark is considered as a heavy quark, and the ansatz $c(x,\mu_0)=0$
with $\mu_0=m$ is assumed as the initial condition for calculating $c(x,\mu_F)$
at higher factorization scales $\mu_F>\mu_0$.
This is usually referred to as the radiatively-generated-charm approach.
In the global analyses cited above, this ansatz implies that the charm parton
does not have any independent degrees of freedom in the parton parameter
space, {\it i.e.},
$c(x,\mu_F)$ is perturbatively determined by the gluon and light-quark parton
parameters.

However, a purely perturbative treatment of the heavy quarks might not be
adequate.
This is even more true for the charm quark with mass $m\approx1.5$~GeV, which
is not much larger than typical hadronic scales of a few hundred MeV.
In fact, there exists the possibility, that $c(x,\mu_0)\ne0$.
Actually, for many years, non-perturbative models exist that give non-zero
predictions for $c(x,\mu_0)$ at some initial factorization scales
$\mu_0=\mathcal{O}(m)$ \cite{Brodsky:1980pb,Navarra:1995rq}.
The model of Ref.~\cite{Brodsky:1980pb} was also analyzed in the framework of
the FFNS \cite{Hoffmann:1983ah}.
In this connection, the notion intrinsic charm (IC) has become customary.
The IC models of Refs.~\cite{Brodsky:1980pb,Navarra:1995rq} and a third one,
which will be specified in a later section, were put to a stringent test by
extending the recent CTEQ6.5 global analysis \cite{Tung:2006tb} so that the
charm sector has its own independent degrees of freedom at the initial
factorization scale $\mu_0=m$ \cite{Pumplin:2007wg}.
As a natural extension of the CTEQ6.5 analysis \cite{Tung:2006tb}, Pumplin
{\it et al.}\ \cite{Pumplin:2007wg} determined the range of magnitude of IC
that is consistent with an up-to-date global analysis of hard-scattering data
with various assumptions on the shape of the $x$ distribution of IC at a low
factorization scale.

The authors of Ref.~\cite{Pumplin:2007wg} found the IC of the light-cone
models \cite{Brodsky:1980pb,Navarra:1995rq} to be compatible with the global
data sample for magnitudes ranging from zero up to three times of what had
been estimated in more model-dependent investigations.
In these models, there can be a large enhancement of $c(x,\mu_F)$ at $x>0.1$,
relative to previous analyses that have no IC.
The enhancement persists to rather large scales $\mu_F$, up to 100~GeV.
Therefore, these nonzero IC contributions can have an important effect on
charm-initiated processes at HERA, RHIC, the Tevatron, and the LHC.
Of course, at hadron colliders, the charm production cross section would be
sensitive to $c(x,\mu_F)$ at $x>0.1$ only at sufficiently large values of
$x_T=2p_T/\sqrt{s}$.
For example, the cross section data at the Tevatron \cite{Acosta:2003ax} only
cover the range $p_T<25$~GeV, so that $x_T<0.025$, which is too small to be
sensitive to IC contributions at $x>0.1$.
Only at RHIC, where $\sqrt{s}$ is a factor of 10 smaller, we may expect
sufficient sensitivity.
At given values of $\sqrt{s}$ and $p_T$, the sensitivity to $c(x,\mu_F)$ at
large values of $x$ could be further enhanced by performing measurements at
large values of $|y|$, {\it i.e.}\ in the extreme forward or backward regions
\cite{Goncalves:2008sw}.

The obvious data that could provide us with information on the IC
contribution, would be those on the charm structure function $F_2^c$ measured
at HERA.
But these data are already used in the global analysis \cite{Pumplin:2007wg}.
Unfortunately, they do not have a significant effect because of the rather
large experimental errors and because the data are mostly located at small
values of $x$.
A comparison of such data with results from earlier CTEQ global analyses may
be found in Ref.~\cite{Thompson:2007mx}.

After these general remarks, it is clear that it is interesting to study cross
sections of inclusive charmed-meson production at RHIC ($\sqrt{s}=200$~GeV)
and the Tevatron ($\sqrt{s}=1.96$~TeV) more closely and to establish those
kinematic regions which are sensitive to the IC component of the proton.
For this purpose, we adopt the GM-VFNS approach outlined in 
Ref.~\cite{Kniehl:2005st,Kniehl:2004fy}.
We first compare the $D^0$, $D^+$, and $D^{*+}$ production data from
Ref.~\cite{Acosta:2003ax} to NLO predictions evaluated with the respective FF
sets \cite{Kneesch:2007ey} so as to test the latter (with or without IC).
Then, we present results for the $D^0$ production cross section under Tevatron
and RHIC experimental conditions assuming the six IC scenarios introduced in
Ref.~\cite{Pumplin:2007wg}.

The outline of this paper is as follows.
In Sec.~\ref{sec:two}, we briefly review the FFs determined in
Ref.~\cite{Kneesch:2007ey}.
In Sec.~\ref{sec:three}, we describe the three IC models investigated in
Ref.~\cite{Pumplin:2007wg} as much as is needed to understand our final
results.
In Sec.~\ref{sec:four}, we present our NLO predictions for inclusive $X_c$
production in $p\overline{p}$ and $pp$ collisions.
Section~\ref{sec:five} contains a summary and an outlook.

\section{Fragmentation functions for charmed mesons}
\label{sec:two}

For our calculation of the differential cross section
$\mathrm{d}^2\sigma/(\mathrm{d}p_T\,\mathrm{d}y)$ of
$p+p(\overline{p})\to X_c+X$, where $X_c=D^0,D^+,D^{*+}$ and $X$ stands for
the residual final state, a crucial ingredient is the non-perturbative FFs for
the transitions $a\to X_c$, where 
$a=g,u,\overline{u},d,\overline{d},s,\overline{s},c,\overline{c}$.
For $X_c=D^{*+}$, such FFs were extracted at LO and NLO in the modified
minimal-subtraction ($\overline{\mathrm{MS}}$) factorization scheme with
$n_f=5$ massless quark flavors several years ago \cite{Binnewies:1997xq} from
the distributions $\mathrm{d}\sigma/\mathrm{d}x$ in scaled energy
$x=2E/\sqrt{s}$ of the cross sections of $e^++e^-\to D^{*+}+X$ measured by the
ALEPH \cite{Barate:1999bg} and OPAL \cite{Ackerstaff:1997ki} Collaborations at
LEP1.
Two of us \cite{Kniehl:2005de} extended the analysis of
Ref.~\cite{Binnewies:1997xq} to include $X_c=D^0,D^+,D_s^+,\Lambda_c^+$ by
fitting appropriate OPAL data \cite{Alexander:1996wy}.
Besides the total $X_c$ yield, which receives contributions from
$Z\to c\overline{c}$ and $Z\to b\overline{b}$ decays as well as from
light-quark and gluon fragmentation, the ALEPH and OPAL Collaborations
separately specified the contributions due to tagged $Z\to b\overline{b}$
events yielding $X_b$ hadrons, which then weakly decay to $X_c$ hadrons.
The contribution due to the fragmentation of primary charm quarks into $X_c$
hadrons approximately corresponds to the difference of these two measured
distributions.

In Refs.~\cite{Binnewies:1997xq,Kniehl:2005de}, the starting point $\mu_0$
for the DGLAP evolution of the $a\to X_c$ FFs in the factorization scale
$\mu_F^\prime$ were taken to be $\mu=2m$ with $m=1.5$~GeV for
$a=g,u,\overline{u},d,\overline{d},s,\overline{s}$ and $\mu=2m_b$ with
$m_b=5$~GeV for $a=b,\overline{b}$.
The FFs for $a=g,u,\overline{u},d,\overline{d},s,\overline{s}$ were assumed to
vanish at $\mu_F^\prime=\mu_0$ and generated through DGLAP evolution to larger
values of $\mu_F^\prime$.
For consistency with the $\overline{\mathrm{MS}}$ prescription for the PDFs,
these fits of the  FFs were repeated for the choice $\mu_0=m,m_b$ in
Ref.~\cite{Kniehl:2006mw}. 
These determinations of $X_c$ FFs
\cite{Kniehl:2006mw,Binnewies:1997xq,Kniehl:2005de} were all based solely on
data from the $Z$-boson resonance.
In these data, the effects of finite quark and hadron masses were greatly
suppressed and could safely be neglected.

The most recent fits for $X_c=D^0,D^+,D^{*+}$ reported in
Ref.~\cite{Kneesch:2007ey} include, beside the OPAL 
\cite{Alexander:1996wy,Ackerstaff:1997ki} and ALEPH \cite{Barate:1999bg} data,
much more precise data from CLEO \cite{Artuso:2004pj} and Belle
\cite{Seuster:2005tr}.
They offered us the opportunity to further constrain the charmed-hadron FFs
and test their scaling violations.
However, the lower c.m.\ energies of the CLEO \cite{Artuso:2004pj} and Belle
\cite{Seuster:2005tr} data necessitates the incorporation of quark and hadron
mass effects, which are then no longer negligible, into the formalism.
The GM-VFNS, which is also the basis of the computation of the cross section
of $p+\overline{p}\to X_c+X$ in Refs.~\cite{Kniehl:2005st,Kniehl:2004fy},
provides the appropriate theoretical framework also for this
(see Ref.~\cite{Kneesch:2007ey} for details).
We adopted the values of $\mu_0$ and $m$ from Ref.~\cite{Kniehl:2006mw}.

In this framework, new FFs for $D^0$, $D^+$, and $D^{*+}$ mesons were
determined through global fits to all available $e^+e^-$ annihilation data,
from Belle \cite{Seuster:2005tr}, CLEO \cite{Artuso:2004pj}, ALEPH
\cite{Barate:1999bg}, and OPAL \cite{Alexander:1996wy,Ackerstaff:1997ki}.
In contrast to the situation at the $Z$-boson resonance, we had to take into
account the effect of electromagnetic initial-state radiation on the Belle and
CLEO data, which distorts the distribution in scaled momentum
$x_p=2p/\sqrt{s}$ of the cross section for continuum production in a
non-negligible way \cite{Kneesch:2007ey}.
We found that the global fits suffer from the fact that the Belle and CLEO
data tend to drive the average of $x$ of the $c\to X_c$ FFs to larger values,
which leads to a somewhat worse description of the ALEPH and OPAL data.
Since the $b\to X_c$ FFs are only indirectly constrained by the Belle and CLEO
data, namely via their contribution to the DGLAP evolution, their forms are
only feebly affected by the inclusion of these data in the fits.
In other words, the $b\to X_c$ FFs are essentially fixed by the ALEPH and OPAL
data alone.
It was found that hadron mass effects are more important than quark mass
effects in the global fits.
In fact, they are indispensable for fitting the lower tails of the $x_p$
distributions from Belle and CLEO

The $z$ distributions of the $c$ and $b$ quark FFs at their starting scales
were assumed to obey the Bowler ansatz \cite{Bowler:1981sb}
\begin{equation}
D_a^{X_c}(z,\mu_0)=Nz^{-(1+\gamma^2)}(1-z)^a\mathrm{e}^{-\gamma^2/z},
\label{eq:bowler}
\end{equation}
with three parameters $N$, $a$, and $\gamma$.
Specifically, the fitting procedure was as follows.
At the scale $\mu_0=m=1.5$~GeV, the $c$-quark FF was taken to be of the form
of Eq.~(\ref{eq:bowler}), while the FFs of the light quarks $q$ ($q=u,d,s$)
and the gluon were set to zero.
Then these FFs were evolved to higher scales $\mu_F^\prime$ using the DGLAP
equations with $n_f=4$ active quark flavors and
$\Lambda_{\overline{\rm MS}}^{(4)}=321$~MeV.
When the scale $\mu_F^\prime$ reached the threshold value
$\mu_F^\prime=m_b=5.0$~GeV, the bottom flavor was activated and its FF was
introduced in the Bowler form of Eq.~(\ref{eq:bowler}).
The evolution to higher scales $\mu_F^\prime$ was then performed with $n_f=5$
and $\Lambda_{\overline{\rm MS}}^{(5)}=221$~MeV.
The values of the Bowler parameters $N_c$, $a_c$, and $\gamma_c$ for
$c\to X_c$ and $N_b$, $a_b$, and $\gamma_b$ for $b\to X_c$ thus obtained may
be found in Tables 1, 2, and 3 of Ref.~\cite{Kneesch:2007ey} for the $D^0$,
$D^+$, and $D^{*+}$ mesons, respectively, together with the achieved values of
$\chi^2$ per d.o.f.\ ($\overline{\chi ^2}$).
The $\overline{\chi ^2}$ values differ somewhat for the three $D$-meson
species, but they are all acceptable.
The smallest $\overline{\chi ^2}$ value was obtained for the $D^+$ meson.

\section{PDFs with intrinsic charm}
\label{sec:three}

In a recent paper \cite{Pumplin:2007wg}, Pumplin {\it et al.}\ extended the
up-to-date CTEQ6.5 global analysis \cite{Tung:2006tb} to include a charm
sector with nonzero $c(x,\mu_0)$ at the initial factorization scale $\mu_0=m$.
For this purpose, they considered three scenarios.
The first two scenarios are based on the light-cone Fock-space picture of
nucleon structure formulated in more detail by Brodsky \cite{Brodsky:2004er}.
In this picture, IC is mainly present at large momentum fraction $x$, because
states containing heavy quarks are suppressed according to their off-shell
distance, which is proportional to $(p_T^2+m^2)/x$.
Thus, components with large $m$ appear preferentially at large $x$.
A wide variety of light-cone models predict similar shapes in $x$, as has been
shown recently \cite{Pumplin:2005yf}.
Specifically, the light-cone models considered in Ref.~\cite{Pumplin:2007wg}
include the original model of Brodsky, Hoyer, Peterson, and Sakai (BHPS)
\cite{Brodsky:1980pb} and the so-called meson-cloud picture
\cite{Navarra:1995rq}, in which the IC arises from virtual low-mass
meson-plus-baryon components of the proton, such as
$\overline{D}^0\Lambda_c^+$.
The meson-cloud model also predicts a difference between $c(x,\mu_F^\prime)$
and $\overline{c}(x,\mu_F^\prime)$ and so provides an estimate of the possible
charm/anticharm difference.

Unfortunately, the light-cone formalism does not allow us to calculate the
normalization of the $uudc\overline{c}$ components \cite{Pumplin:2005yf},
although estimates of the order of 1\% were reported in the literature
\cite{Navarra:1995rq,Donoghue:1977qp}.
Therefore, in Ref.~\cite{Pumplin:2007wg}, the magnitude of IC is determined by
comparison with data incorporated in the global fit.

The $x$ dependence originating in the BHPS model \cite{Brodsky:1980pb} is
\begin{equation}
c(x,\mu_0)=\overline{c}(x,\mu_0)=Ax^2[6x(1+x)\ln x+(1-x)(1+10x+x^2)].
\label{eq:bhps}
\end{equation}
In Eq.~(\ref{eq:bhps}), $\mu_0=m$ with $m=1.3$~GeV and the normalization
constant $A$ is treated as a free parameter in the fit.
Its magnitude is controlled by the average $c+\overline{c}$ momentum fraction,
\begin{equation}
\langle x\rangle_{c+\overline{c}}=\int_0^1\mathrm{d}x\,x[c(x,\mu_0)
+\overline{c}(x,\mu_0)].
\end{equation}
The exact $x$ dependence predicted by the meson-cloud model cannot be given by
a simple formula.
However, it was shown in Ref.~\cite{Pumplin:2005yf} that the charm
distributions in this model can be very well approximated by
\begin{eqnarray}
c(x,\mu_0)&=&Ax^{1.897}(1-x)^{6.095},
\nonumber\\
\overline{c}(x,\mu_0)&=&\overline{A}x^{2.511}(1-x)^{4.929},
\end{eqnarray}
where the normalization constants $A$ and $\overline{A}$ are constrained by
the quark number sum rule,
\begin{equation}
\int_0^1\mathrm{d}x\,[c(x,\mu_0)-\overline{c}(x,\mu_0)]=0,
\end{equation}
which fixes $A/\overline{A}$.
The overall magnitude of IC is again determined in the global analysis.

As a third scenario, Pumplin {\it et al.}\ \cite{Pumplin:2007wg} studied a
purely phenomenological scenario in which the shape of the charm
distribution is sea-like, {\it i.e.}\ similar to those of the light-flavor
sea quarks, except for an overall mass suppression.
In particular, they assumed that
$c(x,\mu_0)=\overline{c}(x,\mu_0)\propto
\overline{u}(x,\mu_0)+\overline{d}(x,\mu_0)$ 
at $\mu_0=m$.

The $c(x,\mu_0)$ and $\overline{c}(x,\mu_0)$ functions of the BHPS,
meson-cloud, and sea-like models discussed above are used in
Ref.~\cite{Pumplin:2007wg} as input for the general-mass perturbative QCD
evolution as explained in Ref.~\cite{Tung:2006tb}.
Then, the range of the IC magnitude is determined to be consistent with the
global data fit.
The quality of each global fit is measured by $\chi_{\mathrm{global}}^2$,
which is shown in Fig.~1 of Ref.~\cite{Pumplin:2007wg} as a function of
$\langle x\rangle_{c+\overline{c}}$ for the three models.
From this figure, one can see that in the lower range,
$0<\langle x\rangle_{c+\overline{c}}\alt0.01$, $\chi_{\mathrm{global}}^2$
varies very little, {\it i.e.} the fit is very insensitive to
$\langle x\rangle_{c+\overline{c}}$ in this interval.
This means that the global analysis of the hard-scattering data gives no
evidence either for or against IC up to
$\langle x\rangle_{c+\overline{c}}\approx0.01$.
For $\langle x\rangle_{c+\overline{c}}>0.01$, the curves for the three models
in Fig.~1 of Ref.~\cite{Pumplin:2007wg} rise steeply with
$\langle x\rangle_{c+\overline{c}}$.
$\chi_{\mathrm{global}}^2\approx3450$ represents a marginal fit in each model,
beyond which the quality of the fit becomes unacceptable according to the
procedure established in Refs.~\cite{Pumplin:2002vw,Tung:2006tb}.
This criterion is based on the fact that one or more of the individual
experiments in the global fit is no longer fitted within the 90\% confidence
level.
For comparison, the fit with no IC yields
$\chi_{\mathrm{global}}^2\approx3330$.
The $\langle x\rangle_{c+\overline{c}}$ distributions of
$\chi_{\mathrm{global}}^2$ achieved in the BHPS, meson-cloud, and sea-like
models reach the marginal-fit limit of 3450 at 2.0\%, 1.8\%, and 2.4\%,
respectively.
The authors of Ref.~\cite{Pumplin:2007wg} studied also the typical, more
moderate $\langle x\rangle_{c+\overline{c}}$ values 0.57\%, 0.96\%, and
1.1\% in the same order.
The $x$ dependences of $c(x,\mu_F)$ and $\overline{c}(x,\mu_F)$ obtained in
the three models are shown for the two values of
$\langle x\rangle_{c+\overline{c}}$ mentioned above and for various values of
$\mu_F$ in Figs.~2--4 of Ref.~\cite{Pumplin:2007wg} and are compared there with
the zero-IC result of the CTEQ6.5 fit.
From these figures one notices that, in the two light-cone models, IC leads to
an enhancement of $c(x,\mu_F)$ and $\overline{c}(x,\mu_F)$ at $x>0.1$ relative
to the PDF analysis with zero IC, while the deviation is small for 
$x<0.1$.
In the sea-like model, there is also a significant enhancement of $c(x,\mu_F)$
and $\overline{c}(x,\mu_F)$ relative to the zero-IC PDF.
In this case, the enhancement is spread more broadly in $x$, roughly over the
region $0.01<x<0.50$.

The six PDF sets constructed with IC, corresponding to the three models with
two values of $\langle x\rangle_{c+\overline{c}}$ each, which are designated
by CTEQ6.5C in Ref.~\cite{Pumplin:2007wg}, are available together with the
corresponding zero-IC PDF set via the LHAPDF standard \cite{LHAPDF} as seven
members, which we denote as CTEQ6.5Cn with $n=0,1,2,3,4,5,6$.
Specifically, $n=0$ stands for zero IC,
$n=1,2$ for the BHPS model with
$\langle x\rangle_{c+\overline{c}}=0.57\%,2.0\%$, 
$n=3,4$ for the meson-cloud model with 
$\langle x\rangle_{c+\overline{c}}=0.96\%,1.8\%$, and
$n=5,6$ for the sea-like model with 
$\langle x\rangle_{c+\overline{c}}=1.1\%,2.4\%$.

\section{Comparison with CDF data and predictions for RHIC}
\label{sec:four}

\subsection{Predictions for CDF}

In this section, we present our predictions for the differential cross section
$\mathrm{d}^2\sigma/(\mathrm{d}p_T\,\mathrm{d}y)$ of $p+\overline{p}\to X_c+X$
with $X_c=D^0,D^+,D^{*+}$ at NLO in the GM-VFNS.
The calculation proceeds as outlined in Ref.~\cite{Kniehl:2004fy} with the
modifications explained in Sec.~\ref{sec:two}.

The NLO cross section consists of three classes of contributions.
\begin{enumerate}
\item Class (i) contains all the partonic subprocesses with a
$c,\overline{c}\to X_c$ transition in the final state that have only light
partons ($g,q,\overline{q}$) in the initial state, the possible pairings being
$gg$, $gq$, $g\bar q$, and $q\bar q$.
\item Class (ii) contains all the partonic subprocesses with a
$c,\overline{c}\to X_c$ transition in the final state that also have $c$ or
$\overline{c}$ quarks in the initial state, the possible pairings being $gc$,
$g\overline{c}$, $qc$, $q\overline{c}$, $\overline{q}c$,
$\overline{q}\overline{c}$, and $c\overline{c}$.
\item Class (iii) contains all the partonic subprocesses with a
$g,q,\bar q\to X_c$ transition in the final state.
\end{enumerate}
The contributions of classes (ii) and (iii) are calculated in the ZM-VFNS
using the hard-scattering cross sections usually used for the inclusive
production of light mesons \cite{Aversa:1988fv}.
The light-quark fragmentation contributions are negligible.
However, gluon fragmentation contributes significantly, as was first noticed
in our previous work \cite{Kniehl:2005st}.

As was shown in Ref.~\cite{Kniehl:2004fy}, the $m$ dependence of the
class-(i) contribution is greatly screened in the full cross section because
the latter is dominated by the contributions of classes (ii) and (iii), which
are calculated with $m=0$.
In fact, the bulk of the contribution is due to class (ii), which provides a
handle on the $c(x,\mu_F)$ and $\overline{c}(x,\mu_F)$ and thus on IC.

The CDF data \cite{Acosta:2003ax} come as distributions
$\mathrm{d}\sigma/\mathrm{d}p_T$ at $\sqrt{s}=1.96$~TeV with $y$ integrated
over the range $|y|\le1$.
For each $X_c$ meson, the particle and antiparticle contributions are
averaged.
We work in the GM-VFNS with $n_f=4$ thus excluding $X_c$ hadrons from
$X_b$-hadron decays, which are vetoed in the CDF analysis
\cite{Acosta:2003ax}.
For the comparison with the CDF data, we use set CTEQ6.5 of proton PDFs
\cite{Tung:2006tb} without IC (see Sec.~\ref{sec:three}), and set global-GM of
$X_c$ FFs \cite{Kneesch:2007ey} (see Sec.~\ref{sec:two}).
We adopt the value $\Lambda_{\overline{\mathrm{MS}}}^{(4)}=328$~MeV, which
yields $\alpha_s^{(5)}(M_Z)=0.1181$, from Ref.~\cite{Tung:2006tb} and the
values $m=1.5$~GeV and $m_b=5$~GeV from Ref.~\cite{Kneesch:2007ey}.
There is a slight mismatch of the input parameters $n_f$,
$\Lambda_{\overline{\mathrm{MS}}}^{(4)}$, $m$, and $m_b$, since
Ref.~\cite{Tung:2006tb} uses $n_f=5$, $m=1.3$~GeV and $m_b=4.5$~GeV and
Ref.~\cite{Kneesch:2007ey} uses $n_f=5$ and
$\Lambda_{\overline{\mathrm{MS}}}^{(4)}=321$~MeV.
The difference between the values of $\Lambda_{\overline{\mathrm{MS}}}^{(4)}$
is certainly insignificant.
The different choices for $m$ and $m_b$, which set the flavor thresholds, will
feebly affect the evolution of the strong-coupling constant, the PDFs, and the
FFs.
In order to conservatively estimate the scale uncertainty, we set
$\mu_R=\xi_Rm_T$, $\mu_F=\mu_F^\prime=\xi_Fm_T$, independently vary $\xi_R$
and $\xi_F$ in the range $1/2<\xi_R,\xi_F<2$, and determine the maximum upward
and downward deviations from our default predictions, for $\xi_R=\xi_F=1$.

\begin{figure}[ht]
\begin{center}
\begin{tabular}{ccc}
\parbox{0.33\textwidth}{
\epsfig{file=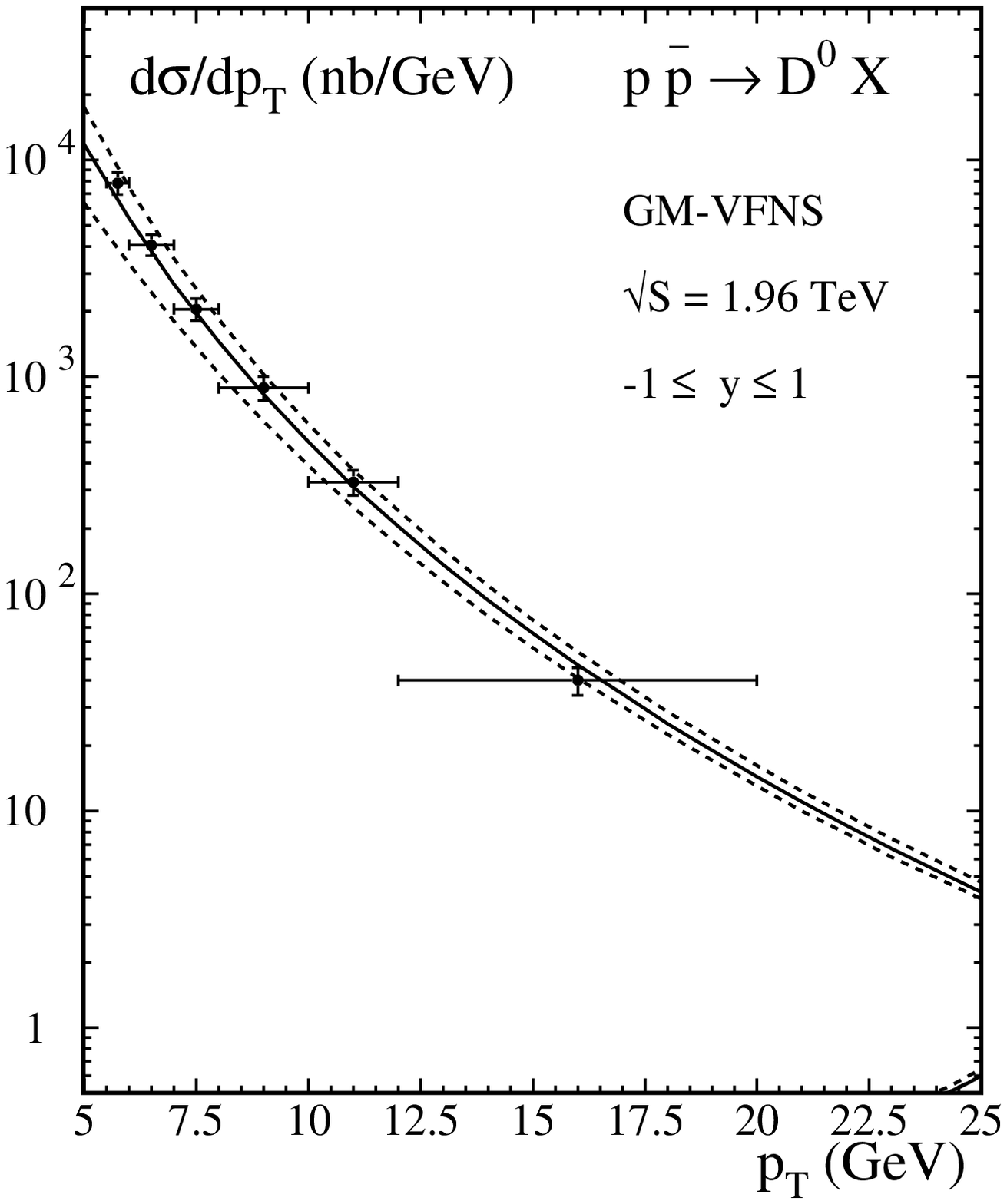,width=0.33\textwidth}
}
&
\parbox{0.33\textwidth}{
\epsfig{file=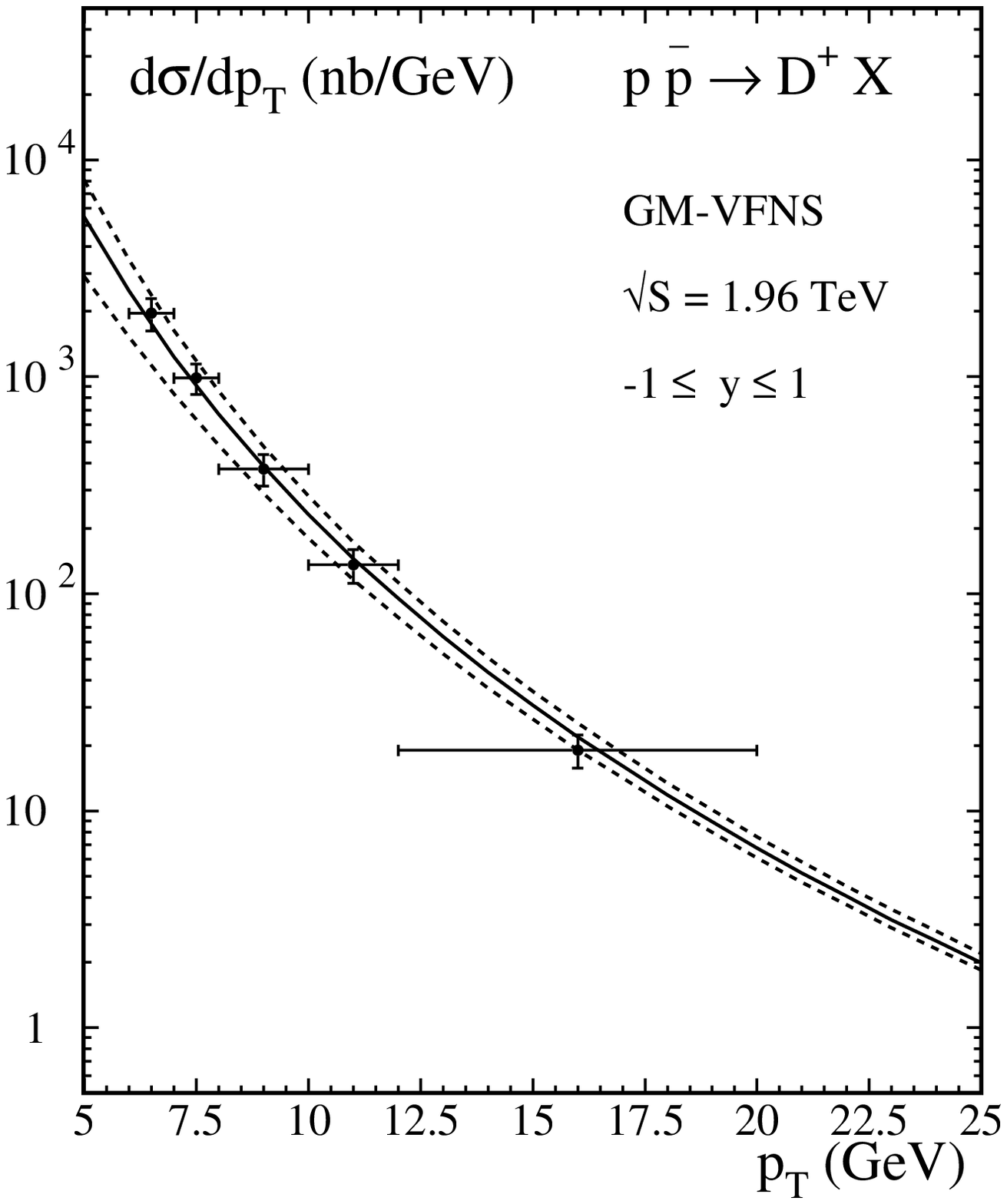,width=0.33\textwidth}
}
&
\parbox{0.33\textwidth}{
\epsfig{file=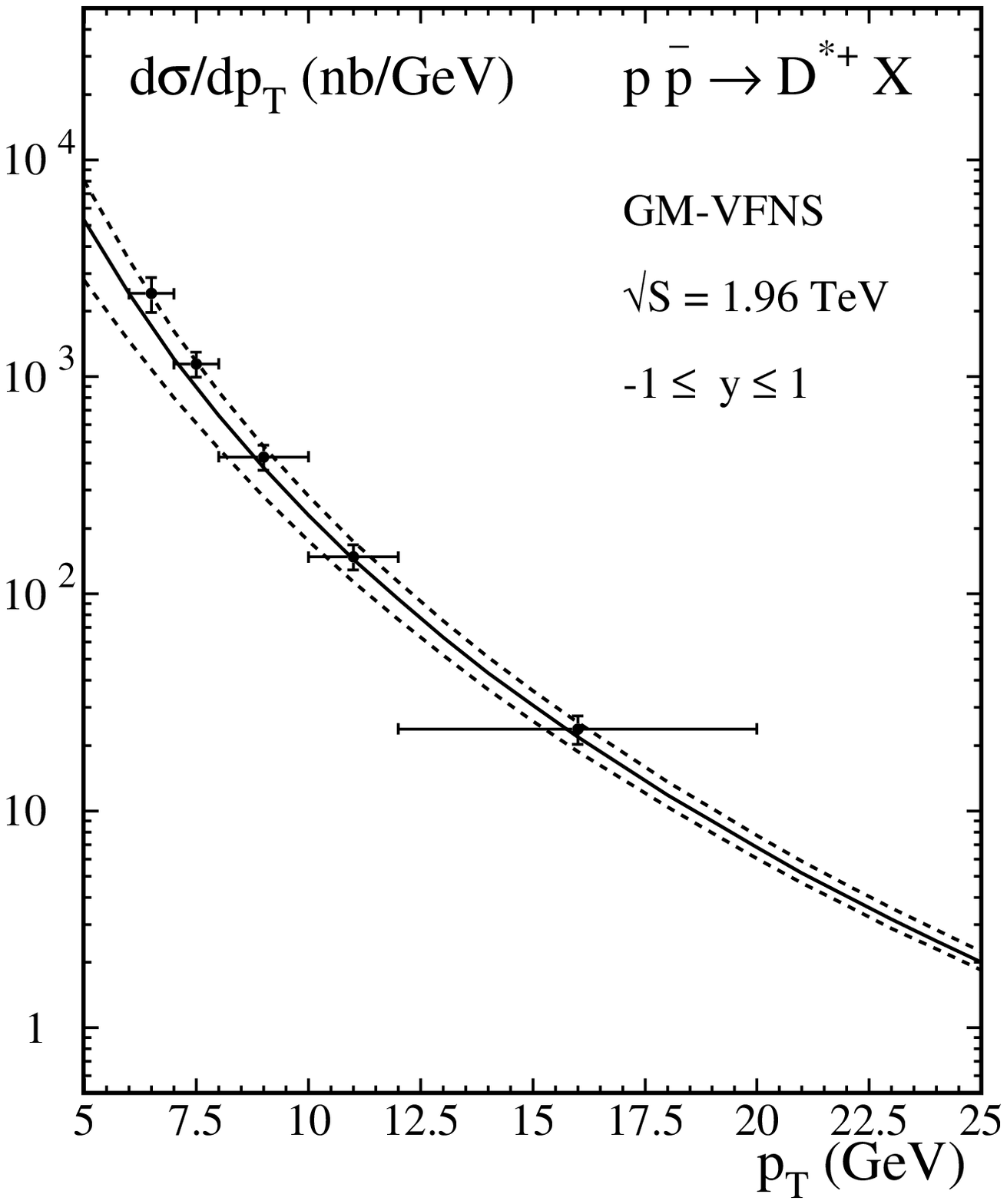,width=0.33\textwidth}
}
\\
(a) & (b) & (c)
\end{tabular}
\end{center}
\caption{\label{fig:one}%
$p_T$ distributions $\mathrm{d}\sigma/\mathrm{d}p_T$ of
$p+\overline{p}\to X_c+X$ with (a) $X_c=D^0$, (b) $X_c=D^+$, and (c)
$X_c=D^{*+}$ for $\sqrt{s}=1.96$~TeV and $|y|<1$ evaluated at NLO in the
GM-VFNS using the FFs of Ref.~\cite{Kneesch:2007ey} in comparison with
experimental data from CDF \cite{Acosta:2003ax}.
The solid lines represents the default predictions, for $\xi_R=\xi_F=1$, and
the dashed lines indicate the maximum deviations for independent variations in
the range $1/2<\xi_R,\xi_F<2$.}
\end{figure}
\begin{figure}[ht]
\begin{center}
\begin{tabular}{ccc}
\parbox{0.33\textwidth}{
\epsfig{file=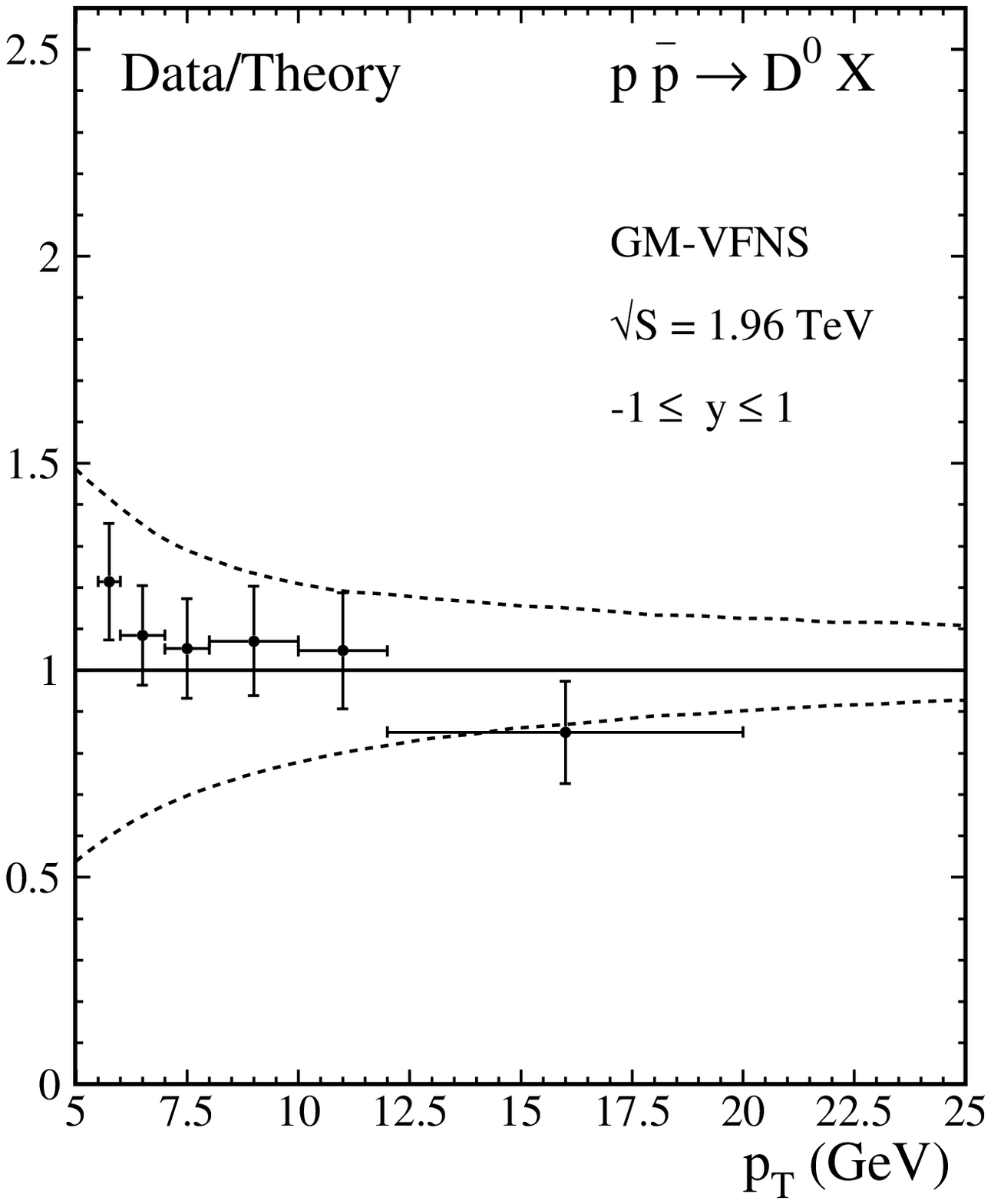,width=0.33\textwidth}
}
&
\parbox{0.33\textwidth}{
\epsfig{file=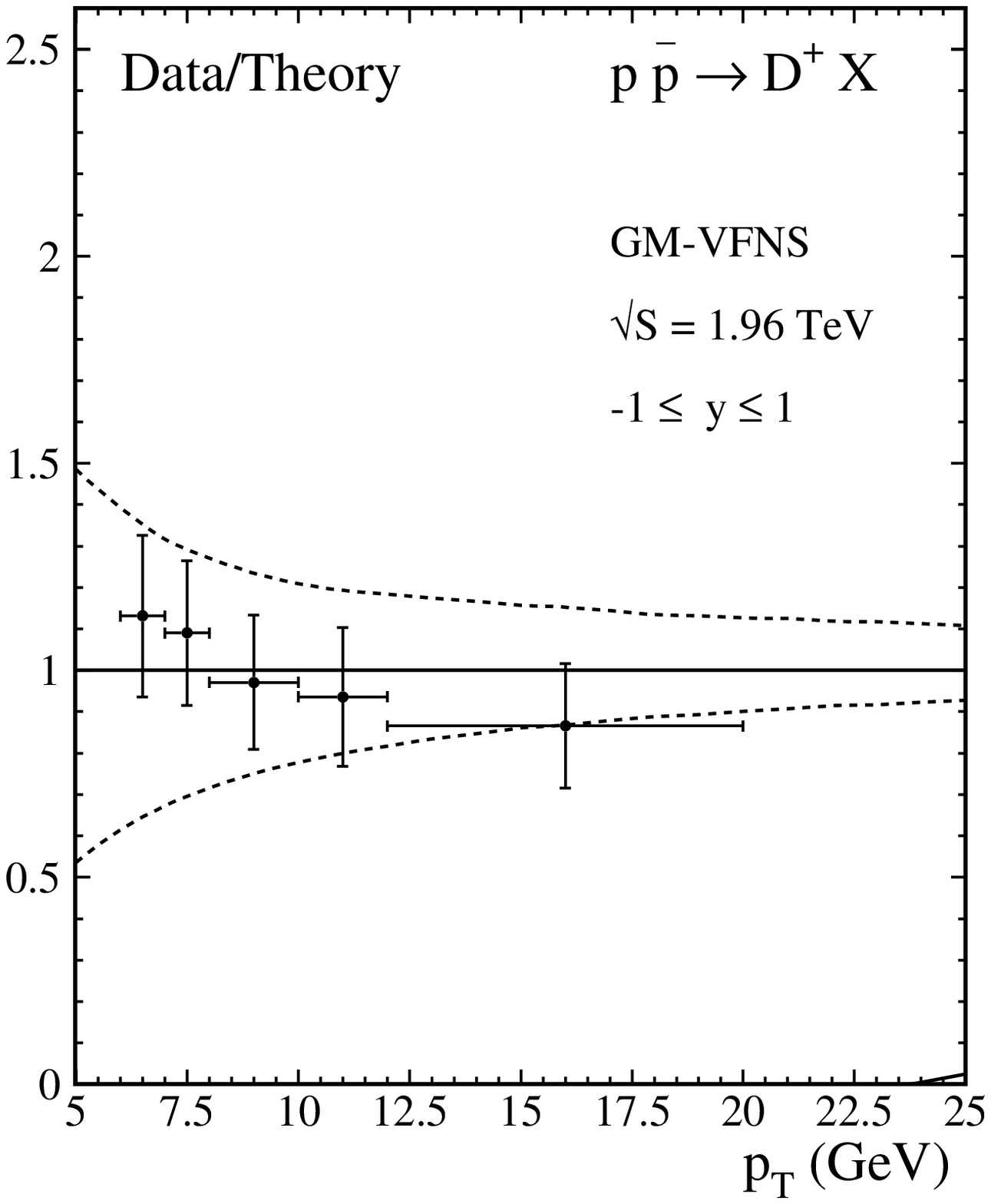,width=0.33\textwidth}
}
&
\parbox{0.33\textwidth}{
\epsfig{file=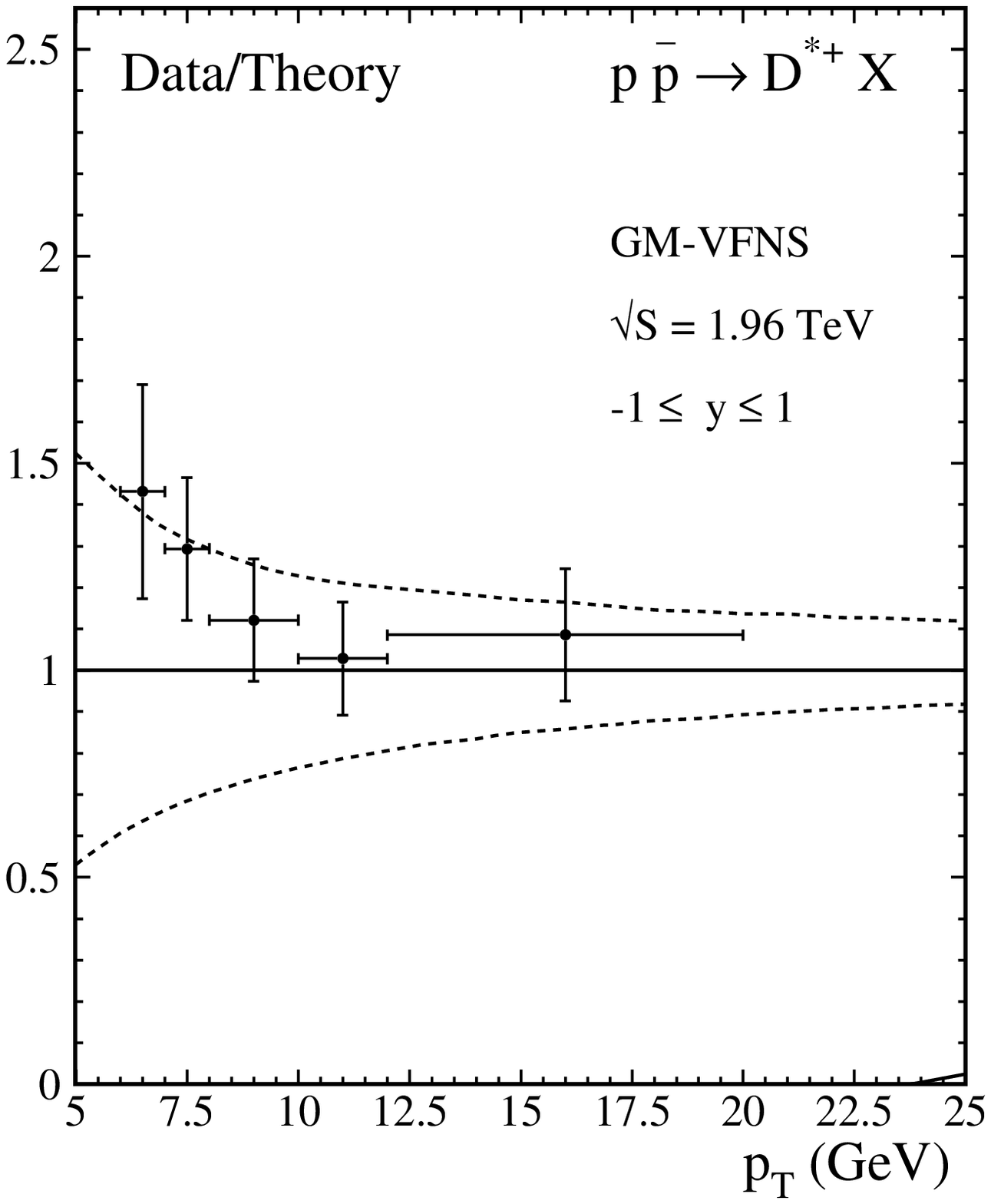,width=0.33\textwidth}
}
\\
(a) & (b) & (c)
\end{tabular}
\end{center}
\caption{\label{fig:two}%
Same as in Figs.~\ref{fig:one}(a)--(c), but normalized to the default
predictions.}
\end{figure}

The theoretical predictions evaluated with the CTEQ6.5 PDFs \cite{Tung:2006tb}
and the new FFs \cite{Kneesch:2007ey} are compared with the CDF
data \cite{Acosta:2003ax} on an absolute scale in Fig.~\ref{fig:one} and in
the data-over-theory representation with respect to the default results in
Fig.~\ref{fig:two}.
The three frames in each figure refer to the $D^0$, $D^+$, and $D^{*+}$
mesons; in each frame, the theoretical uncertainty due to the scale variation
described above is indicated by the dashed lines.
In all cases, we find that the agreement with the data is improved relative to
our previous analysis \cite{Kniehl:2005st}, as may be seen by comparing
Figs.~\ref{fig:one} and \ref{fig:two} with Figs.~1 and 2 in 
Ref.~\cite{Kniehl:2005st}.
In fact, the $D^0$ data, except for the smallest-$p_T$ point, and the $D^+$
data agree with our default predictions within the experimental errors.
This is also true for the $D^{*+}$ data, except for the two small-$p_T$ points.
This improvement may be attributed to the advancement in our knowledge of
charmed-meson FFs \cite{Kneesch:2007ey}, which now also includes detailed
information from the $B$ factories, while the previously used FF set was
solely based on LEP1 data \cite{Kniehl:2006mw}.

\begin{figure}[ht]
\begin{center}
\epsfig{file=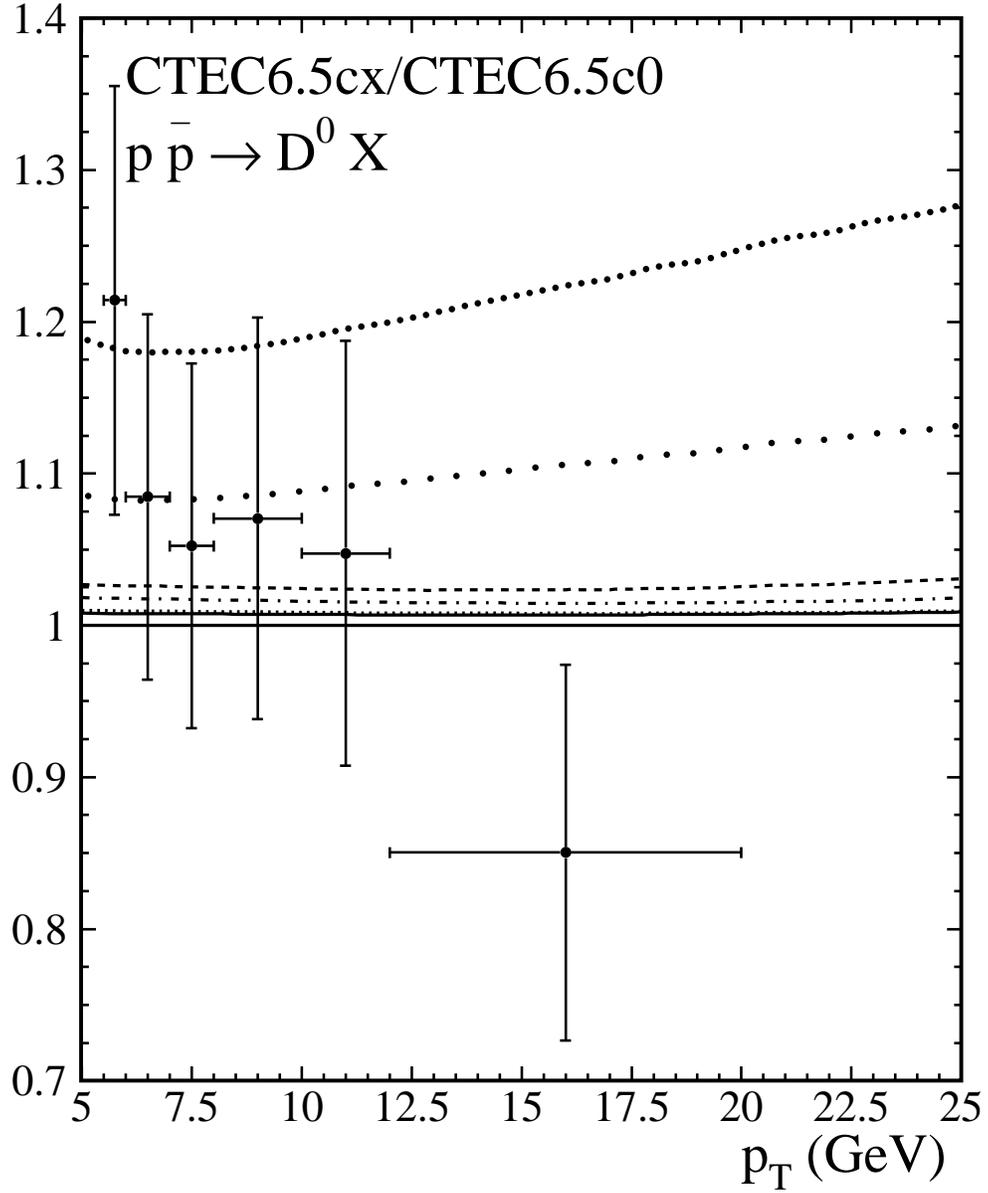,width=0.95\textwidth}
\end{center}
\caption{\label{fig:three}%
Same as in Fig.~\ref{fig:two}(a), but including besides the default prediction
those evaluated with the IC parameterizations from Ref.~\cite{Pumplin:2007wg}
for $n=1$ (solid line), 2 (dashed line), 3 (densely dotted line), 4
(dot-dashed line), 5 (scarcely dotted line), 6 (dotted line).}
\end{figure}
In the following, we take the point of view that the FFs for the $X_c$ mesons
are sufficiently well known and ask the question whether the CDF data can
discriminate between the various CTEQ6.5 PDF sets endowed with IC.
For brevity, we only present our results for the case of the $D^0$ meson,
which yields the largest cross section; the results for the $D^+$ and $D^{*+}$
mesons are very similar.
Specifically, we repeat the calculation of the central prediction in
Fig.~\ref{fig:one}(a) in turn with PDF sets CTEQ6.5Cn for $n=1,\ldots,6$ and
normalize the outcome to the default prediction with zero IC of
Fig.~\ref{fig:one}(a).
The results are shown in Fig.~\ref{fig:three}.
We observe that the ratios for $n=1,2,3,4$ lie very close to unity, the
largest deviation being 2\%.
Only the ratios for $n=5,6$ significantly deviate from unity.
In fact, for $n=5$ the ratio is around 1.1, and for $n=6$ it ranges between
1.17 and 1.27 in the considered $p_T$ range.
This finding is easy to understand.
The $x$ values dominantly contributing to the production cross section at the
Tevatron are typically rather small, {\it e.g.}\ $x<x_T\approx0.025$ for
$p_T=25$~GeV.
In this $x$ range, IC is greatly suppressed for $n=1,2,3,4$, while the IC for
$n=5,6$ is significant for $x>10^{-3}$.
The IC for $n=6$, with $\langle x\rangle_{c+\overline{c}}=0.024$, is more
pronounced than that for $n=5$, with
$\langle x\rangle_{c+\overline{c}}=0.011$, which explains the hierarchy of the
respective ratios in Fig.~\ref{fig:three}.
We conclude from Fig.~\ref{fig:three} that the size of the IC enhancement is
comparable to the errors on the CDF data only for $n=5,6$.
In fact, the two large-$p_T$ data points from CDF tend to disfavor the
sea-like IC implemented for $n=6$, although a firm statement would be
premature.
On the other hand, the effects due the IC of the BHPS ($n=1,2$) or meson-cloud
($n=3,4$) models are too feeble to be resolved by the presently available CDF
data.
In order to obtain a handle on these types of IC, one needs data at
considerably larger values of $p_T$, where the deviations from the zero-IC
prediction are sufficiently large.
Of course, the cross sections will be much smaller there and hard to measure
with sufficient accuracy.

\begin{figure}[ht]
\begin{center}
\begin{tabular}{cc}
\parbox{0.45\textwidth}{
\epsfig{file=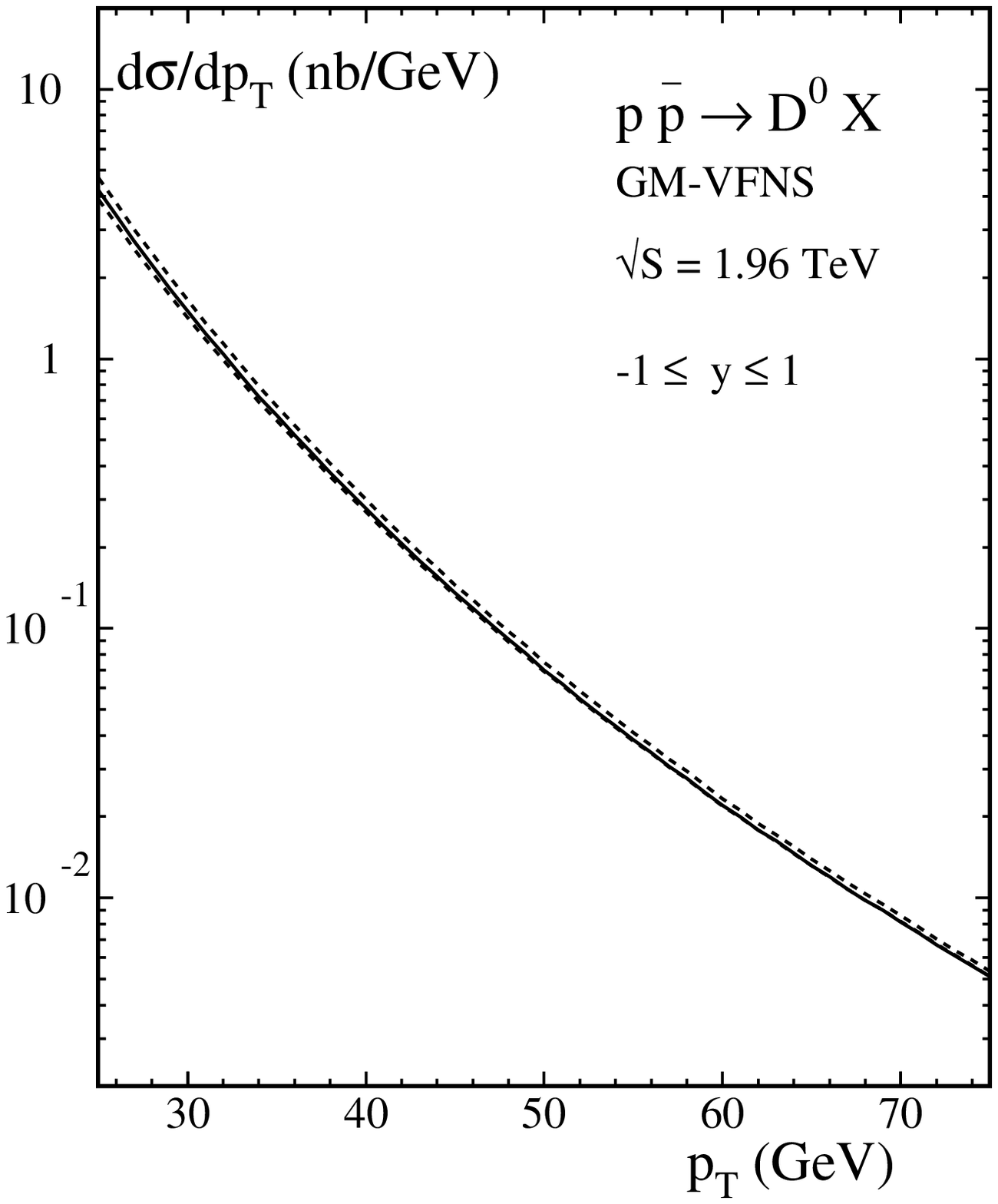,width=0.45\textwidth}
}
&
\parbox{0.45\textwidth}{
\epsfig{file=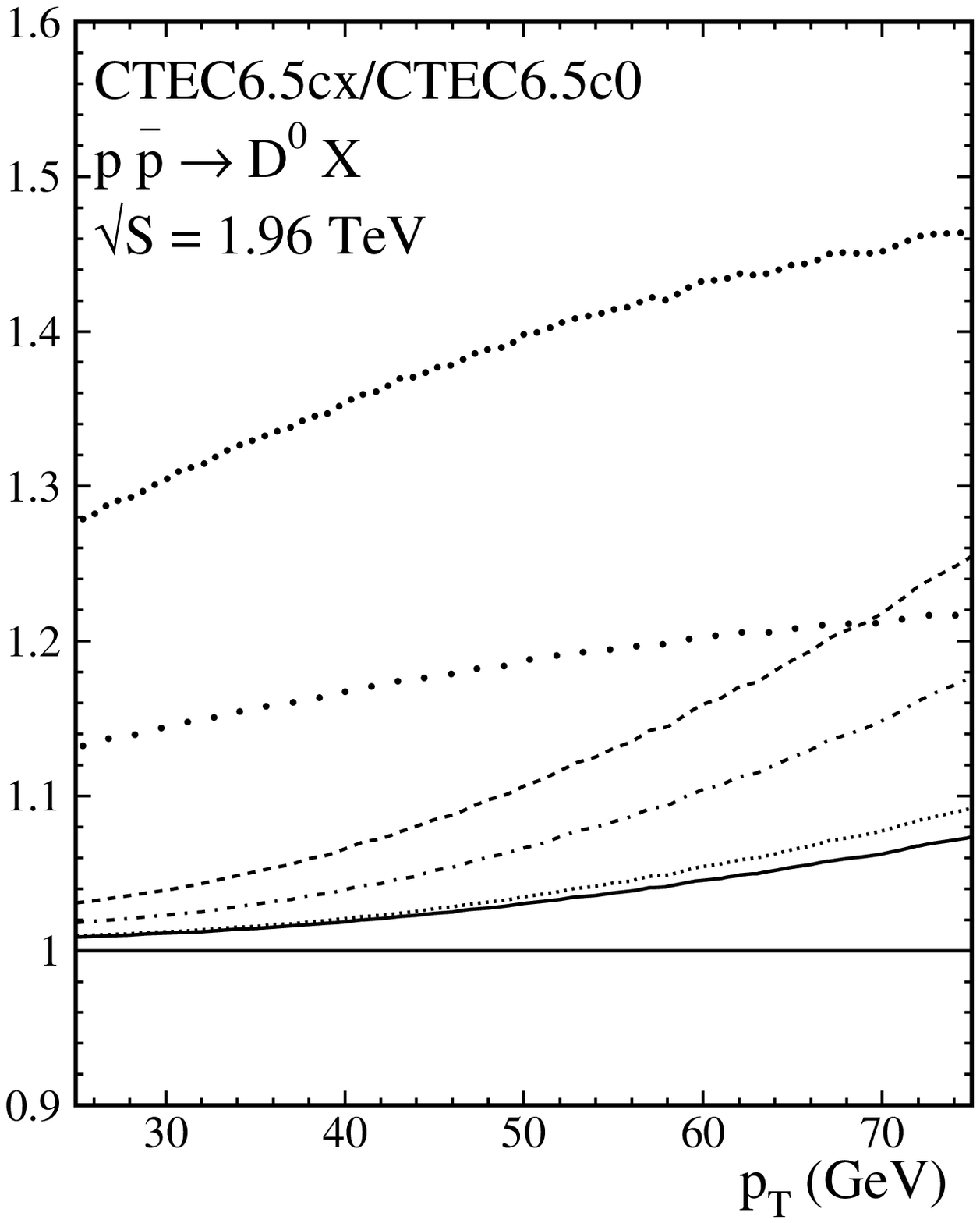,width=0.45\textwidth}
}
\\
(a) & (b)
\end{tabular}
\end{center}
\caption{\label{fig:four}%
Same as in Figs.~\ref{fig:one}(a) and \ref{fig:three}, but for
25~GeV${}<p_T<75$~GeV.}
\end{figure}
However, one should keep in mind that the CDF analysis of
Ref.~\cite{Acosta:2003ax} is merely based on 5.8~pb$^{-1}$ of data recorded in
February and March 2002.
Since then the integrated luminosity of run II has increased by more than a
factor of 1000, exceeding 6~fb$^{-1}$ as it does.
Therefore, we conclude this section by exploring the $p_T$ range beyond
25~GeV, up to 75~GeV, assuming the data to be taken in the central tracking
region, $|y|<1$, as before.
The respective extensions of Figs.~\ref{fig:one}(a) and \ref{fig:three} are
presented as Figs.~\ref{fig:four}(a) and (b).
From Fig.~\ref{fig:four}(a), we read off that the cross section decreases from
5 to $5\times10^{-3}$~nb/GeV as $p_T$ runs from 25 to 75~GeV, and that the
theoretical uncertainty ranges from small to negligible.
Comparing Fig.~\ref{fig:four}(b) with Fig.~\ref{fig:three}, we observe that,
in the BHPS and meson-cloud models, the sensitivity to IC is dramatically
increased as $p_T$ runs from 25 to 75~GeV, while it reaches some level of
saturation in the sea-like model.
We conclude that the measurements of the cross section distributions
$\mathrm{d}\sigma/\mathrm{d}p_T$ of $p\overline{p}\to X_c+X$ based on the
full data sample to be collected at the Tevatron by the end of run~II would
have the potential to yield useful constraints on IC.

\subsection{Predictions for RHIC}

\begin{figure}[ht]
\begin{center}
\begin{tabular}{ccc}
\parbox{0.33\textwidth}{
\epsfig{file=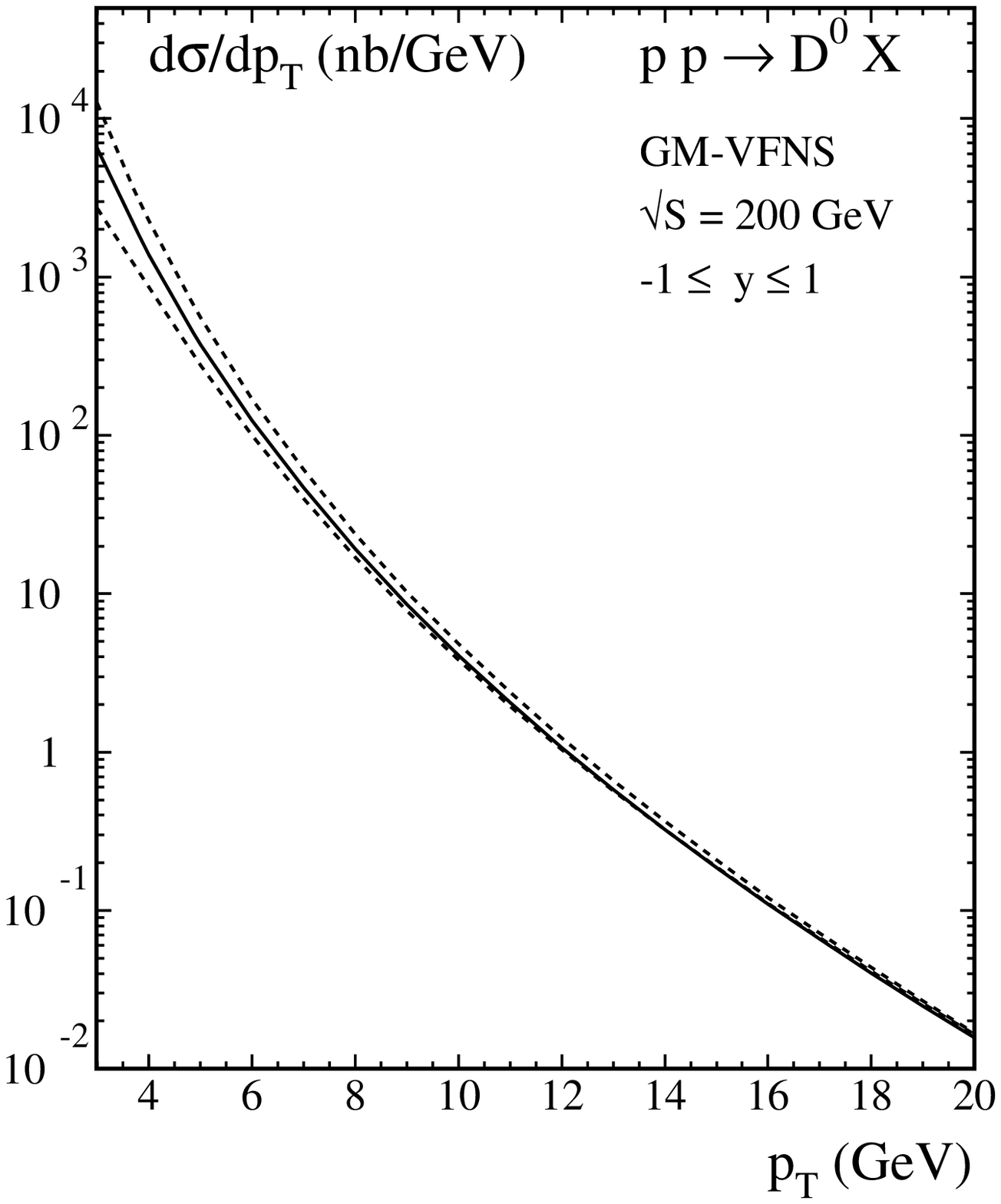,width=0.33\textwidth}
}
&
\parbox{0.33\textwidth}{
\epsfig{file=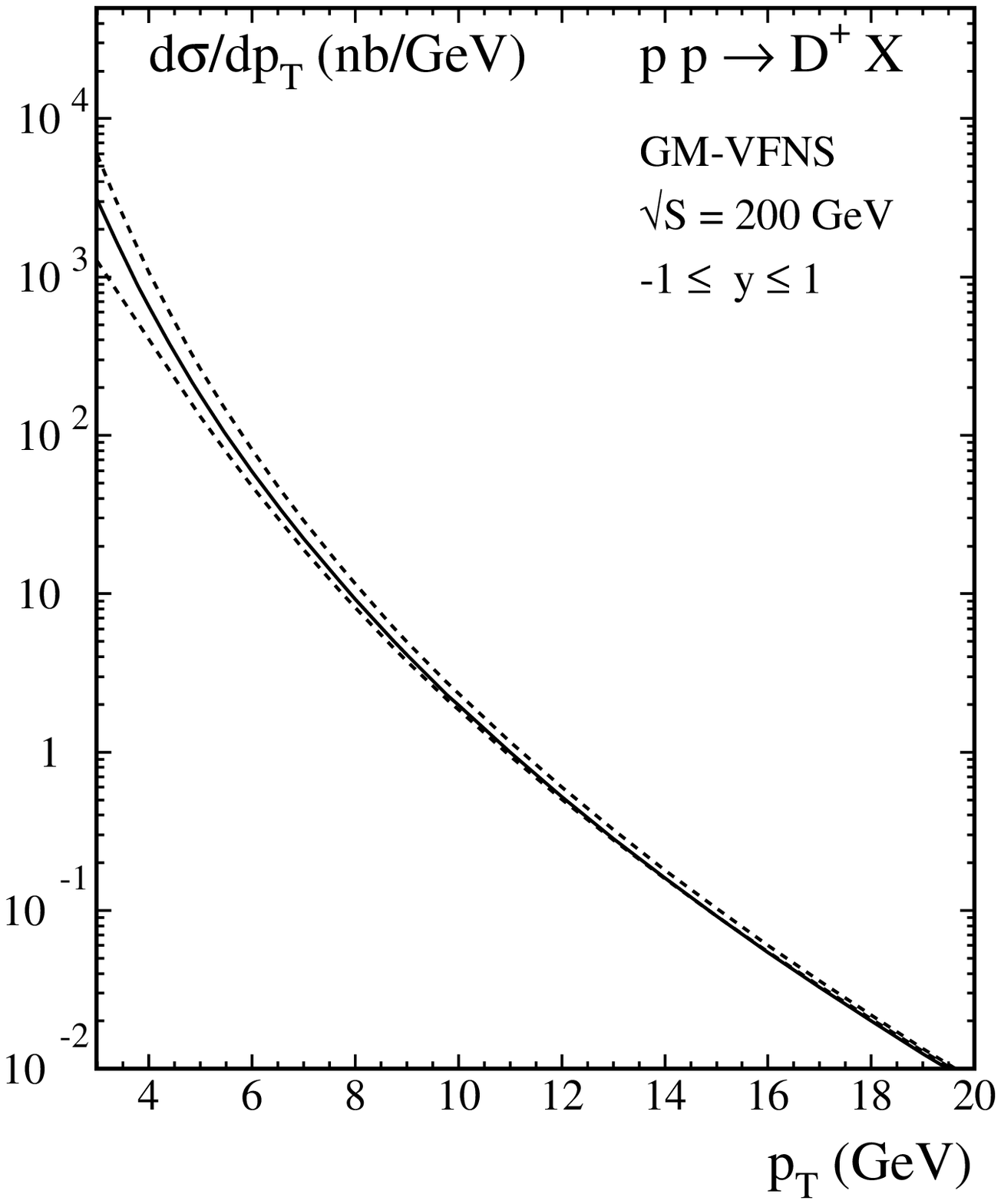,width=0.33\textwidth}
}
&
\parbox{0.33\textwidth}{
\epsfig{file=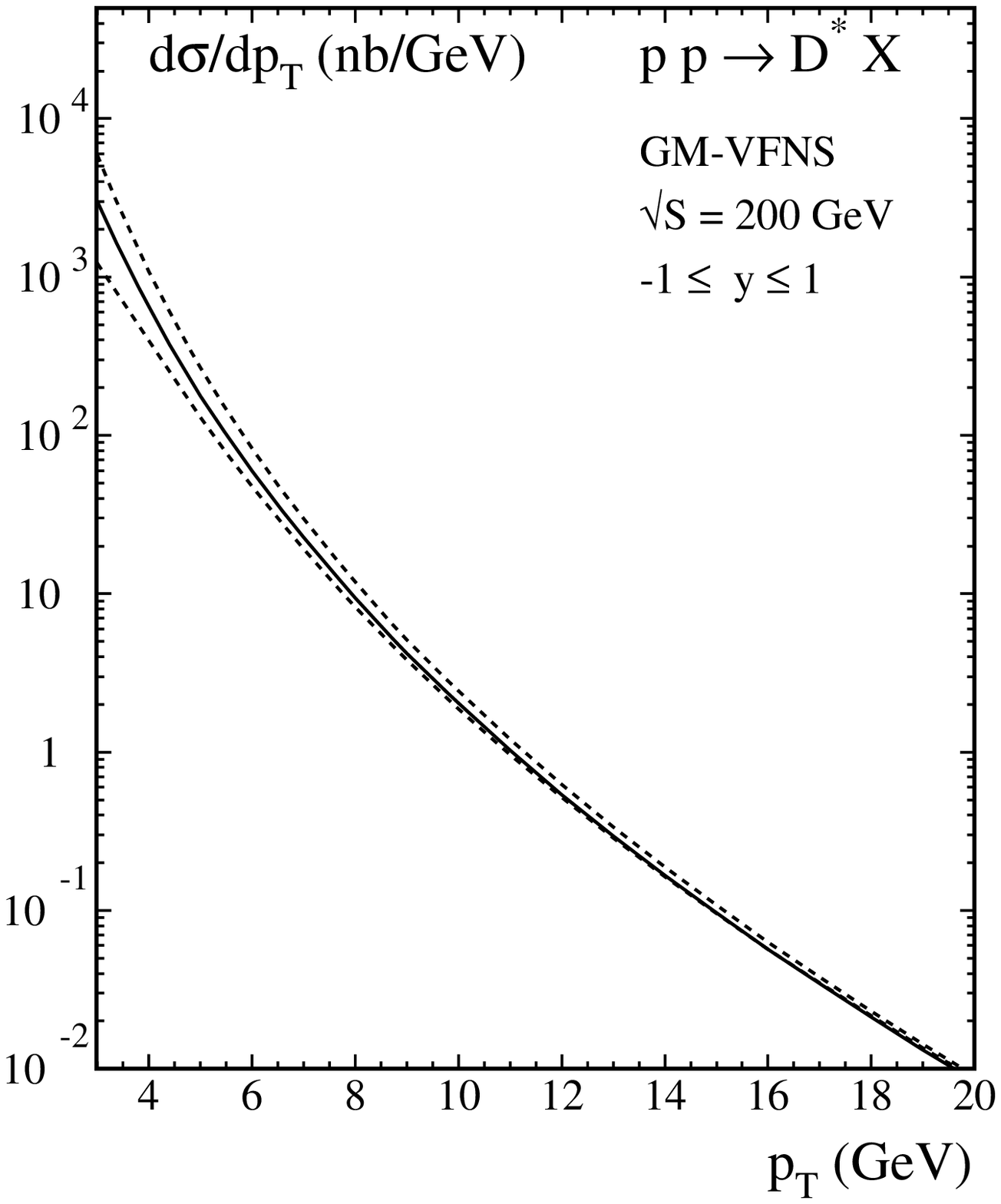,width=0.33\textwidth}
}
\\
(a) & (b) & (c)
\end{tabular}
\end{center}
\caption{\label{fig:five}%
Same as in Fig.~\ref{fig:one}, but for $pp$ collisions with
$\sqrt s=200$~GeV.}
\end{figure}
Instead of measuring the cross sections at much larger values of $p_T$ with
the CDF and D0 detectors at the Tevatron, it may be advantageous to lower the
c.m.\ energy to decrease $x_T$ in the considered $p_T$ range.
A collider with smaller c.m.\ energy is in operation, namely RHIC with
$\sqrt s=200$~GeV, where the cross section distributions
$\mathrm{d}\sigma/\mathrm{d}p_T$ of $pp\to X_c+X$ could be measured.
So far, only the STAR Collaboration has published $X_c$ production data, namely
for $d\mathrm{Au}\to D^0+X$ with $p_T<2.5$~GeV \cite{Adams:2004fc}.
We encourage STAR and PHENIX to also study $pp\to D^0+X$ at larger values of
$p_T$.

The calculations of $\mathrm{d}^2\sigma/(\mathrm{d}p_T\,\mathrm{d}y)$ for
$pp\to X_c+X$ at RHIC are completely analogous to the Tevatron case.
We only need to replace the antiproton PDFs by proton PDFs and put
$\sqrt s=200$~GeV.
We again integrate $y$ over the range $|y|<1$.
In Fig.~\ref{fig:five}, we show $\mathrm{d}\sigma/\mathrm{d}p_T$ as a function
of $p_T$ in the range 3~GeV${}<p_T<20$~GeV for $X_c=D^0,D^+,D^{*+}$.
The results for $D^+$ and $D^{*+}$ are very similar, while that for $D^0$ has
a similar shape, but differs in normalization.
In fact, the $D^{*+}$ to $D^0$ cross section ratio ranges from 0.46 to 0.52 in
the considered $p_T$ range.
As in the case of the Tevatron, the cross sections of $pp\to X_c+X$ at RHIC
rapidly fall with increasing value of $p_T$, by the factor $2.4\times10^{-6}$
as $p_T$ runs from 3~GeV to 20~GeV.
Thus, it might be difficult to measure these cross sections in the upper part
of the considered $p_T$ range.

\begin{figure}[ht]
\begin{center}
\epsfig{file=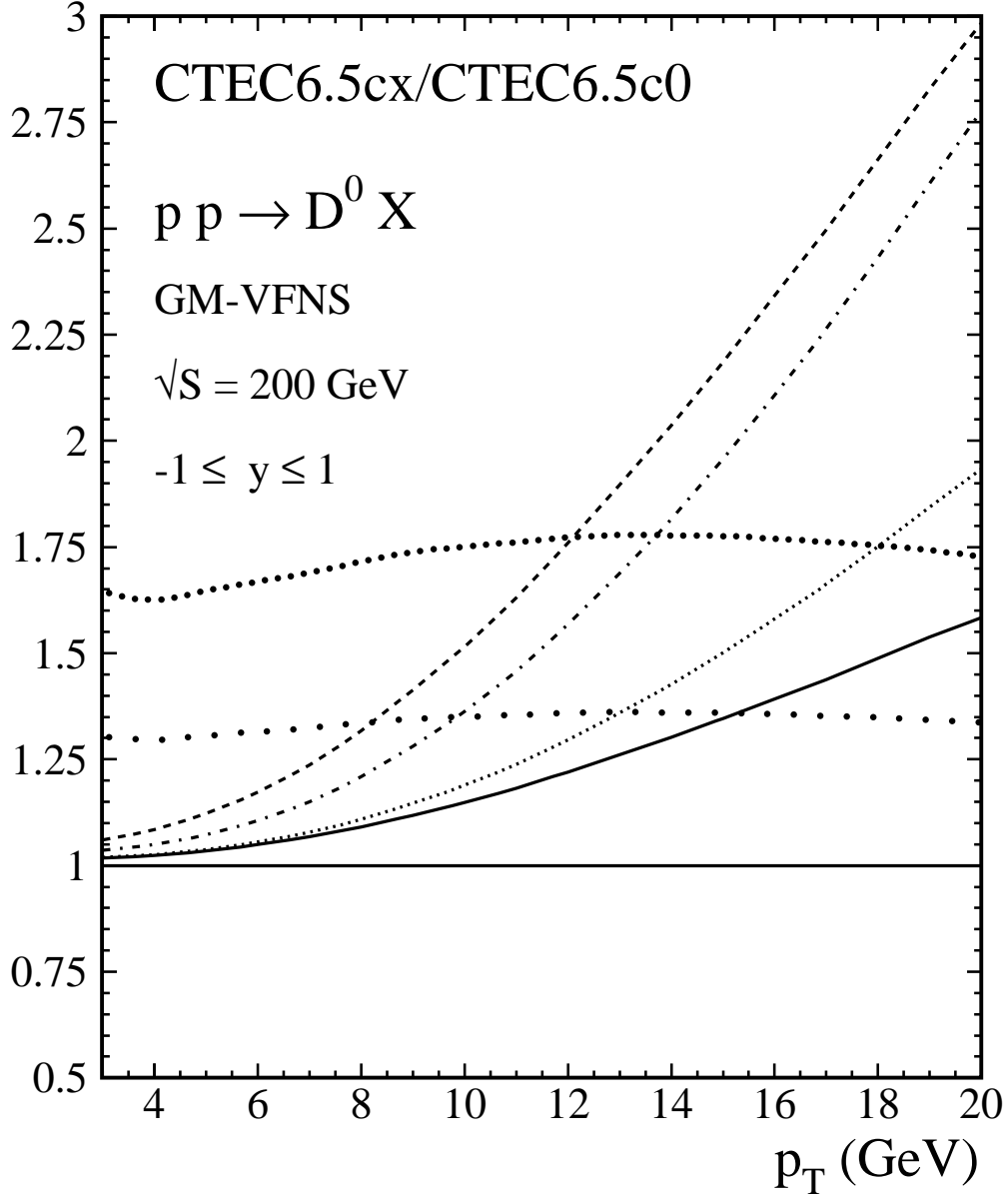,width=0.95\textwidth}
\end{center}
\caption{\label{fig:six}%
Same as in Fig.~\ref{fig:three}, but for $pp$ collisions with
$\sqrt s=200$~GeV.}
\end{figure}
We now repeat the analysis for the $D^0$ meson in Fig.~\ref{fig:five} using
the various PDF sets with IC, CTEQ6.5Cn with $n=1,2,3,4,5,6$.
The results, normalized to the calculation for $n=0$, are shown in
Fig.~\ref{fig:six}.
Here we observe a pattern familiar from the Tevatron case in
Fig.~\ref{fig:three}, except that now these ratios are much larger.
In fact, the results for $n=2,4$ steeply rise with increasing value of $p_T$,
reaching values of about 3 and 2.8 at $p_T=20$~GeV.
The results for $n=1,3$ exhibit less dramatic rises because the IC is weaker
in these cases.
Comparing Figs.~\ref{fig:three} and \ref{fig:six}, we conclude that RHIC
offers a much higher sensitivity to IC as implemented in CTEQ6.5Cn with
$n=1,\ldots,4$ than the Tevatron.
On the other hand, the results for $n=5,6$ at RHIC are rather similar to those
at the Tevatron, being almost independent of $p_T$.
But also here, the relative deviation from the zero-IC case is stronger for
RHIC.

\begin{figure}[ht]
\begin{center}
\begin{tabular}{cc}
\parbox{0.45\textwidth}{
\epsfig{file=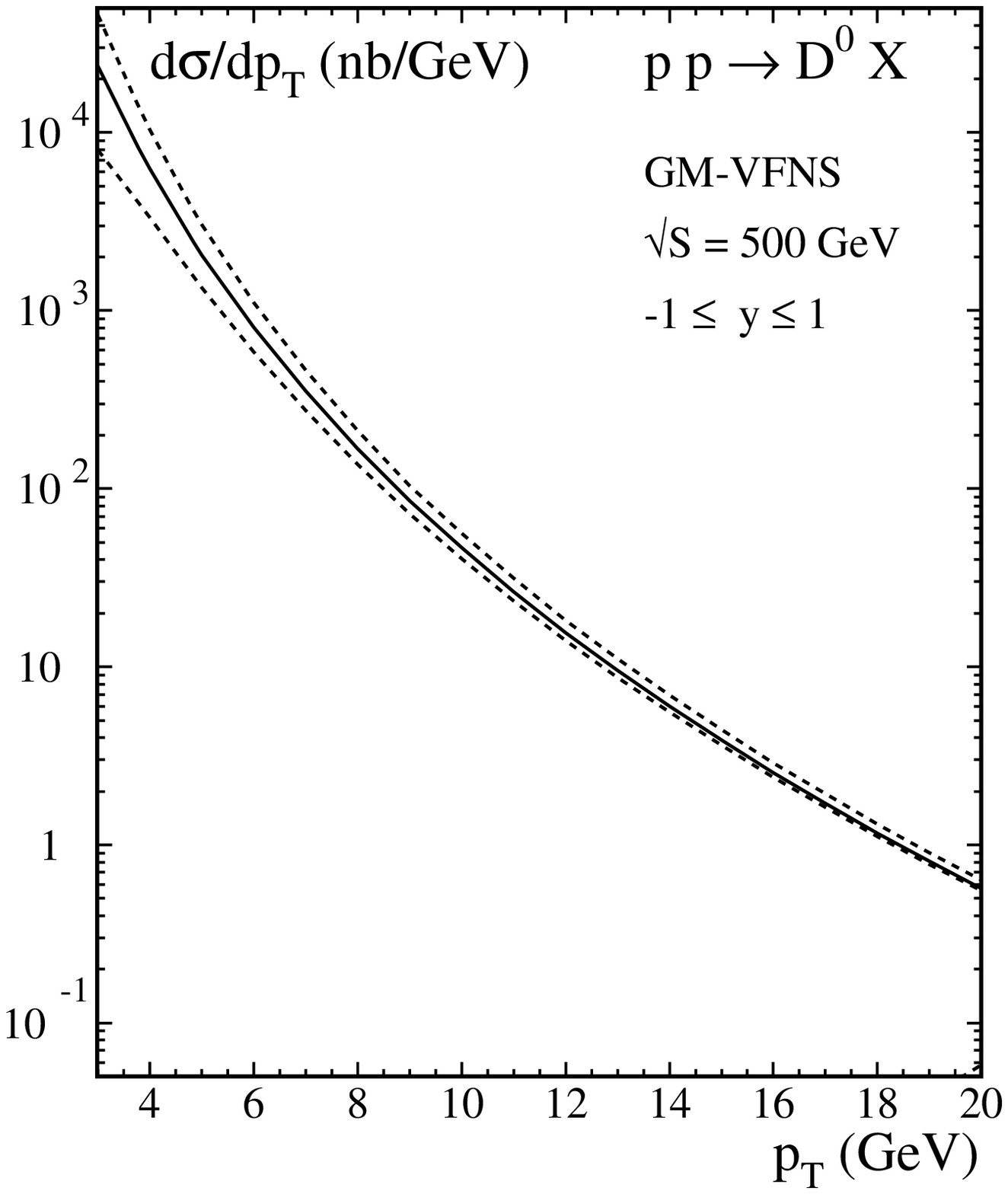,width=0.45\textwidth}
}
&
\parbox{0.45\textwidth}{
\epsfig{file=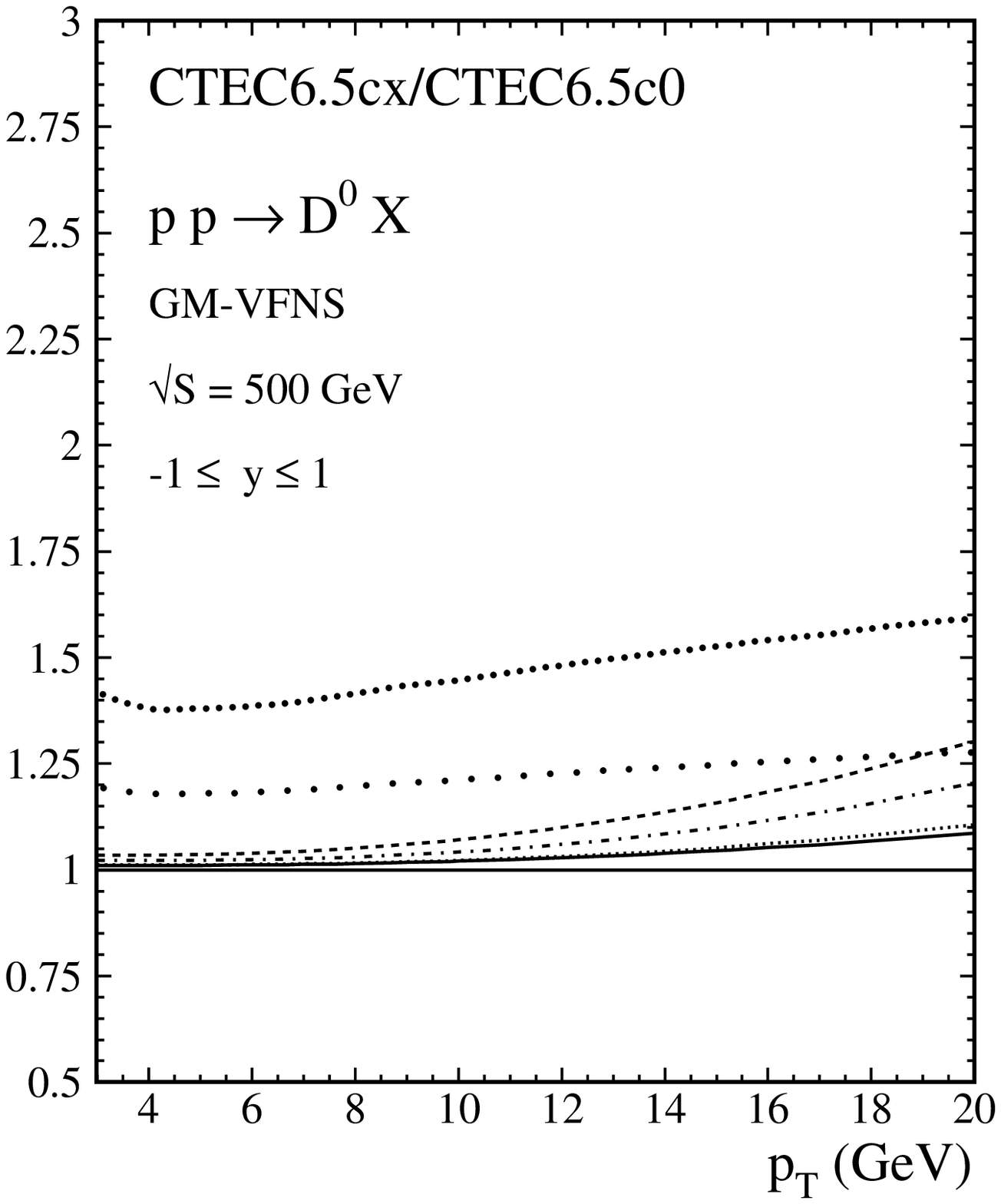,width=0.45\textwidth}
}
\\
(a) & (b)
\end{tabular}
\end{center}
\caption{\label{fig:seven}%
Same as in Figs.~\ref{fig:five}(a) and \ref{fig:six}, but for
$\sqrt s=500$~GeV.}
\end{figure}
Unfortunately, in contrast to the Tevatron, RHIC measurements at
$\sqrt{s}=200$~GeV will be limited by low luminosity.
This key limitation makes the observation of high-$p_T$ $X_c$ production at
this RHIC energy rather unrealistic.
The high-energy $pp$ mode of RHIC, with $\sqrt{s}=500$~GeV, will accrue more
luminosity, perhaps 200--400~pb$^{-1}$.
However, this will happen at the expense of lowering the accessible $x$ values
and, thus, of reducing the relative IC effects on the cross sections,
especially for the BHPS and meson-cloud models.
In order to assess the trade-off between higher luminosity and smaller relative
shifts in the cross sections, we repeat the $D^0$-meson analysis for the
500~GeV mode of RHIC.
The results are shown in Figs.~\ref{fig:seven}(a) and (b), which should be
compared with Figs.~\ref{fig:five}(a) and \ref{fig:six}, respectively.
We observe that, as one passes from 200~GeV to 500~GeV, the cross section is
increased by a factor of 3.6 (36) at $p_T=3$~GeV (20~GeV) in normalization,
while its relative shifts due to IC are greatly reduced for the BHPS and
meson-cloud models, as expected.  

\section{Conclusions}
\label{sec:five}

In this paper, we updated and improved our previous analysis
\cite{Kniehl:2005st} of charmed-meson inclusive hadroproduction at NLO in the
GM-VFNS by using as input the non-perturbative FFs extracted from a global
analysis of $e^+e^-$ annihilation data from CESR \cite{Artuso:2004pj}, KEKB
\cite{Seuster:2005tr}, and LEP1
\cite{Alexander:1996wy,Ackerstaff:1997ki,Barate:1999bg} in the very same
scheme \cite{Kneesch:2007ey}.
In fact, this led to a significantly better description of the $p_T$
distributions of the $D^0$, $D^+$, and $D^{*+}$ mesons measured at the
Tevatron \cite{Acosta:2003ax}, as becomes evident by comparing
Figs.~\ref{fig:one} and \ref{fig:two} with Figs.~1 and 2 in 
Ref.~\cite{Kniehl:2005st}.

Encouraged by this finding, we then investigated how much room there is for
the incorporation of IC inside the colliding hadrons.
Specifically, we adopted six IC parameterizations \cite{Pumplin:2007wg}, which
are based on the BHPS \cite{Brodsky:1980pb}, meson-cloud
\cite{Navarra:1995rq}, and sea-like \cite{Pumplin:2007wg} models, implemented
with two different values of $\langle x\rangle_{c+\overline{c}}$ each.
For definiteness, we focused on the $D^0$ meson.
In the case of the Tevatron, we found that the BHPS and meson-cloud models
yield insignificant deviations from the zero-IC predictions, while the shift
produced by the sea-like model is comparable to the experimental error and
tends to worsen the agreement between theory and experiment.
However, the experimental errors in Ref.~\cite{Acosta:2003ax} are still too
sizeable to rule out this model as implemented in Ref.~\cite{Pumplin:2007wg}.
This is likely to change once the full data sample of run~II is exploited.

Since IC typically receives a large fraction $x$ of momentum from the parent
hadron and the kinematical upper bound on $x$ scales with $x_T=2p_T/\sqrt{s}$,
the sensitivity to IC may be enhanced by lowering the cm.\ energy $\sqrt{s}$.
This is a strong motivation for studying the inclusive production of charmed
mesons with large values of $p_T$ in $pp$ collisions at RHIC, currently
operated at $\sqrt{s}=200$~GeV.
In fact, we found that all three IC models predict sizeable enhancements of
the $p_T$ distributions of $D^0$ mesons at RHIC, by up to 75\% at
$p_T=10$~GeV.
We, therefore, encourage our colleagues at RHIC to perform a dedicated study
of charmed-meson inclusive production in the high-$p_T$ regime.
In a future high-energy $pp$-collision mode of RHIC, with $\sqrt{s}=500$~GeV,
which is to accrue more luminosity, the cross sections would be increased,
while the relative shifts due to IC would be considerably smaller for the
BHPS and meson-cloud models.

\section*{Acknowledgment}

This work was supported in part by the German Federal Ministry
for Education and Research BMBF through Grant No.\ 05~HT6GUA, by the German
Research Foundation DFG through Grant No.\ KN~365/7--1, and by the Helmholtz
Association HGF through Grant No.\ Ha~101.

\end{document}